\documentclass[]{raa}            

\usepackage{graphicx,times}             
\usepackage{natbib}
\usepackage{amssymb,amsmath}
\usepackage{multirow}
\bibpunct{(}{)}{;}{a}{}{,}

\usepackage{hyperref}
\hypersetup{colorlinks = true, linkcolor = green, anchorcolor = red, citecolor = blue, filecolor = red, pagecolor = red, urlcolor = red}

\begin{document}

   \title{Calibrating photometric redshift measurements with the Multi-channel Imager (MCI) of the China Space Station Telescope (CSST)
}

   \volnopage{Vol.0 (20xx) No.0, 000--000}      
   \setcounter{page}{1}          

   \author{Ye Cao
      \inst{1,2}
   \and Yan Gong
      \inst{1,3*}
   \and Zhen-Ya Zheng
      \inst{4,5}
   \and Chun Xu
      \inst{5,6}      
   }
 \institute{Key Laboratory of Space Astronomy and Technology, National Astronomical Observatories,Chinese Academy of Sciences, Beijing 100012, China; {\it Email: gongyan@bao.ac.cn}\\
         \and
         School of Astronomy and Space Sciences, University of Chinese Academy of Sciences, Beijing 100049, China;\\
          \and
         Science Center for China Space Station Telescope, National Astronomical Observatories, Chinese Academy of Sciences, 20A Datun Road, Beijing 100101, China \\
         \and
         CAS Key Laboratory for Research in Galaxies and Cosmology, Shanghai Astronomical Observatory, Shanghai 200030, China;\\         
          \and
         Division of Optical Astronomical Technologies, Shanghai Astronomical Observatory, Shanghai 200030, China;\\
         \and
         Joint Center of SHAO and SITP for Infrared Astronomical Instrumentation, Shanghai Astronomical Observatory, Shanghai 200030, China;\\
   }
   
\abstract{
The China Space Station Telescope (CSST) photometric survey aims to perform a high spatial resolution ($\sim0.15''$) photometric imaging for the targets that cover a large sky area ($\sim$17,500 ${\rm deg^2}$) and wide wavelength range (from NUV to NIR). It expects to explore the properties of dark matter, dark energy, and other important cosmological and astronomical areas. In this work, we evaluate whether the filter design of the Multi-channel Imager (MCI), one of the five instruments of the CSST, can provide accurate photometric redshift (photo-$z$) measurements with its nine medium-band filters to meet the relevant scientific objectives. We generate the mock data based on the COSMOS photometric redshift catalog with astrophysical and instrumental effects. The application of upper limit information of low signal-to-noise ratio (SNR) data is adopted in the estimation of photo-$z$. We investigate the dependency of photo-$z$ accuracy on the filter parameters, such as band position and width. We find that the current MCI filter design can achieve good photo-$z$ measurements with accuracy $\sigma_z\simeq 0.017$ and outlier fraction $f_{\rm c}\simeq 2.2\%$. It can effectively improve the photo-$z$ measurements of the main CSST survey using the Survey Camera (SC) to an accuracy $\sigma_z\simeq 0.015$ and outlier fraction $f_{\rm c}\simeq1.5\%$. It indicates that the original MCI filters are proper for the photo-z calibration.
}

   \authorrunning{Y. Cao, et al.}            
   \titlerunning{Photo-z calibration with the CSST-MCI}  

   \maketitle

%
%
\section{Introduction}           
\label{sect:intro}

Photometric wide-field surveys can detect billions of target sources, and obtain the positions of these sources on the celestial sphere and the fluxes in multiple bands. It plays a major role in modern astronomy research. A few current and upcoming large ground- and space-borne telescopes, e.g. the Dark Energy Survey (DES) \citep{DES05}, Sloan Digital Sky Survey (SDSS) \citep{Fukugita96,York00}, Large Synoptic Survey Telescope (LSST) \citep{LSST09,Ivezic19}, Euclid space telescope \citep{Laureijs11} and Hyper Suprime-Cam (HSC) \citep{Aihara18} have set photometric sky survey requirements with large areas and deep fields to ensure that their scientific objectives can be achieved. The information provided by these surveys could solve many important cosmological and astronomical questions, such as the properties of dark energy and dark matter, the formation and evolution of galaxies, the origin of the universe, etc.

Redshift provides distance information for targets on the traditional two-dimensional sky map. The high precision measurements of galaxy redshift are traditionally determined by spectroscopic method. However, spectroscopic redshift measurement is time-consuming and it is difficult to achieve the redshift calibration of all sources in large sky surveys. On the other hand, photometric redshift (photo-$z$) measurement as an efficient method plays an increasingly important role in large sky surveys. Photo-$z$ technology has been developed and improved in recent decades \citep{Salvato19}, and it has been sufficiently useful for most recent cosmological and astronomical researches, e.g. weak gravitational lensing, cosmic large scale structure, etc.

The China Space Station Telescope (CSST) is a 2-meter space telescope operating in the same orbit of the China Manned Space Station. As a major science project of the Space Application System of the China Manned Space Program, CSST is scheduled to be launched around 2024. It is designed to be a large field of view ($\sim1.1\ {\rm deg}^2$), high spatial resolution ($\sim0.15''$ and $0.18''$ at 633 nm for Survey Camera (SC) and Multi-Channel Imager (MCI), respectively) and multi-band (from NUV to NIR) telescope that will simultaneously perform both photometric and slitless grating photometric surveys covering a large sky area of 17,500 ${\rm deg}^{2}$ with its SC in about a decade \citep{Zhan11,Cao18,Gong19}. 
CSST will explore a number of cosmological and astronomical objectives, such as dark mater and dark energy, the cosmic large scale structure, galaxy clusters, galaxy formation and evolution, active galactic nucleus, etc.

In this work, we explore the measurements of photo-$z$, whose accuracy is the basis of many important scientific objectives, especially for the weak gravitational lensing survey. According to previous studies, the sample used in weak gravitational lensing must have an accuracy $\sigma<0.05$ in next generation surveys \citep[e.g.][]{LSST09}, and we expects to improve its photo-$z$ accuracy to achieve $\sigma\simeq0.02$ \citep{Zhan06}. Here, we investigate the photo-$z$ accuracy and its dependence on the filter design parameters of the CSST-MCI. The MCI has three channels covering the same wavelength range as the SC from the NUV to NIR bands, and these channels can work simultaneously (Zheng et al. in Prep.). Three sets of filters, i.e. narrow-, medium-, and wide-band filters, will be installed on the MCI to perform extreme-deep field surveys with a field of view $7.5'\times 7.5'$. The magnitude limit can be stacked to a depth of 29-30 AB mag in three channels. It will study the formation and evolution of high-$z$ galaxies, properties of dark matter and dark energy, and also can be used to calibrate the photo-$z$ measurements with its nine medium-band filters for the CSST-SC surveys.

There are mainly two methods for estimating photo-$z$, i.e. ``training method" and  ``template fitting method" \citep{Connolly95,Lanzetta96,Brunner97,Fernandez99,Abdalla11,Sanchez14}, which have different advantages. 
The ``training method'' usually requires detecting the spectroscopic redshifts as training samples, e.g. the neutral network software ANNz \citep{Firth03,Collister04}. 
Although this method has high accuracy, currently it only can estimate the photo-$z$ of the bright sources at low redshifts due to the limitation of spectral data as training sample. On the other hand, 
the ``template fitting method'' extracts the photo-$z$ by fitting photometric data with the spectral energy distribution (SED) templates, and the estimated range of redshift is not restricted.
Publicly available codes of this type include Hyperz \citep{Bolzonella00}, BPZ \citep{Benitez00}, LePhare \citep{Arnouts99,Ilbert06}, CIGALE \citep{Boquien19,Yang21}, etc. Following \cite{Cao18}, we use a modified LePhare software considering the flux upper-limit information to explore the photo-$z$ measurements. The machine learning method also can be used in the study as shown in \cite{Zhou21}.

In order to simulate photometric observation as realistic as possible, we generate the mock data based on the COSMOS photometric redshift catalogs (henceforth referred as COSMOS08 and COSMOS15 catalogs)\citep{Capak07,Ilbert09,Laigle16}.  These two catalogs have similar magnitude limit as the CSST, and hence similar galaxy redshift and magnitude distributions \citep{Cao18,Gong19,Zhou21}.
After considering the effects of redshift and extinction, we generate the mock data by convolving SED and filter curves, and then we estimate their photo-$z$ and analyze the results. 
We explore the capability of each MCI medium-band filter in the photo-$z$ measurements, and investigate the filter design by changing the filter position of central wavelength and the band wavelength width. The results can provide good references for the design of the CSST-MCI medium-band filters.

This paper is organized as follows: in Section~\ref{sect:method}, we introduce the definition of the CSST-MCI medium-band filters and the method of generating mock data based on the COSMOS catalog. In Section~\ref{sect:est}, we use a modified LePhare software to estimate the photo-$z$, and then investigate the dependency of photo-$z$ accuracy on the band and filter parameters for the MCI medium-band filters. Finally we summarize the results in Section~\ref{sect:sum}.

\section{Mock data}
\label{sect:method}

\subsection{The definition of the MCI medium-band filters}

\begin{figure}
   \centering
   \includegraphics[width=0.495\textwidth, angle=0]{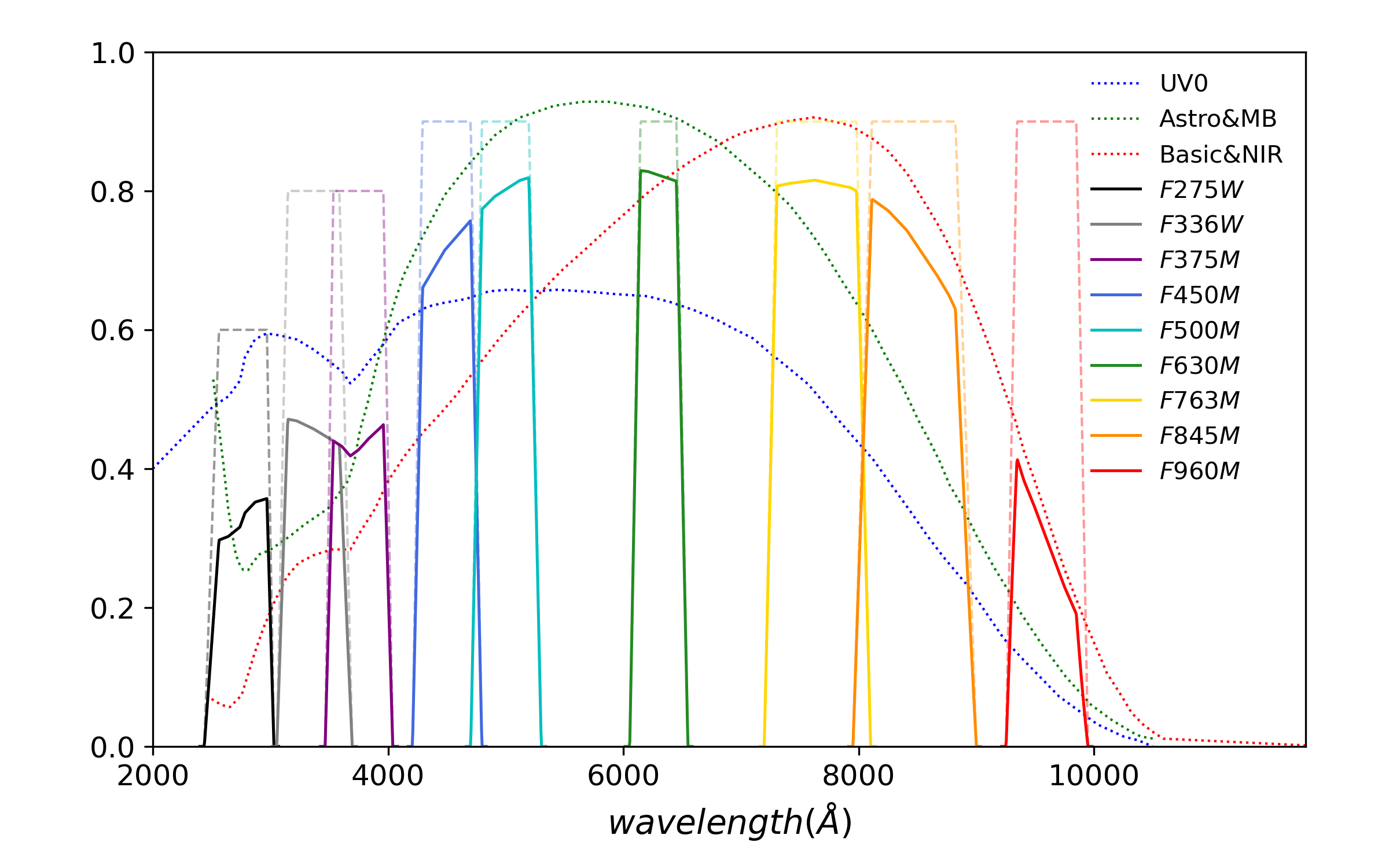}
   \includegraphics[width=0.495\textwidth, angle=0]{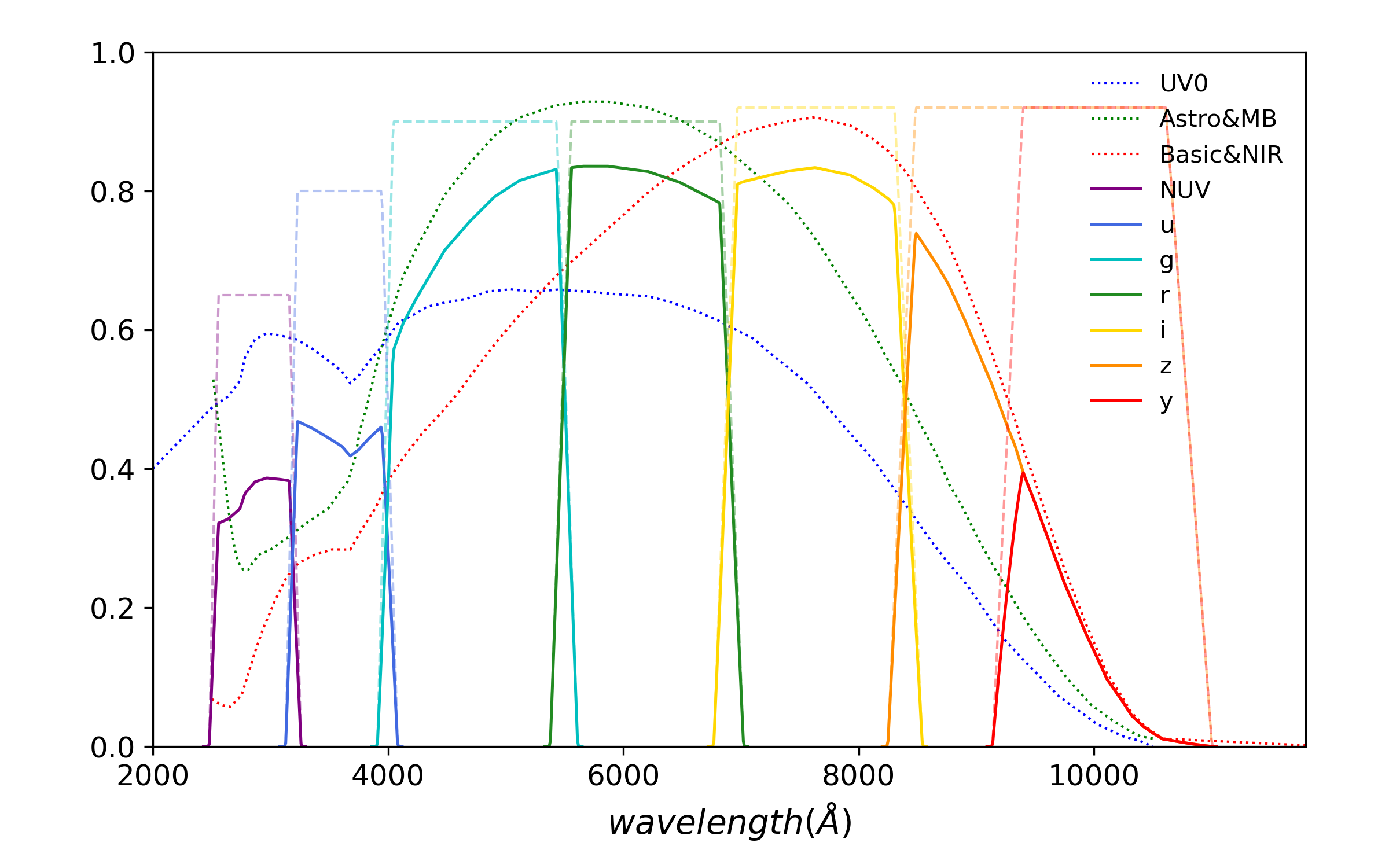}
   \caption{ $Left$: The transmission curves for the nine MCI medium-band filters from NUV to NIR bands, including the $F275W$, $F336W$, $F375M$, $F450M$, $F500M$, $F630M$, $F763M$, $F845M$ and $F960M$.
   $right$: The transmission curves for the seven SC filters, including the $NUV$, $u$, $g$, $r$, $i$, $z$, and $y$ bands. 
   The dotted curves show the detector quantum efficiency, and the dashed lines and the solid lines are the intrinsic transmission and the total transmission by considering detector quantum efficiency, respectively. }
   \label{fig:filters}
\end{figure}

Based on the current instrumental design, the CSST-MCI medium-band filters include 9 filters from near-ultraviolet to near-infrared, including $F275W$, $F336W$, $F375M$, $F450M$, $F500M$, $F630M$, $F763M$, $F845M$ and $F960M$. We set the curve shape of the intrinsic transmission of the filters to trapezoid, and show the transmission curves for the filters that are under test in Figure~\ref{fig:filters}. The left panel show the transmission curves of the MCI medium-band filters, which are used to explore the dependency of photo-$z$ accuracy. The right panel is the transmission curves of the SC broad filters, which are based on \cite{Cao18}. Here the intrinsic transmission and the total transmission including the detector quantum efficiency are presented in dashed and solid lines, respectively. We list the definition of the MCI medium-band filters in Table \ref{tab:filters}. Here we show the mean wavelength $\lambda_{\rm c}$, full width at half maximum (FWHM), and the wavelengths at 50\% of the maximum transmission curve ( left: $\lambda_{\rm L50}$, and right: $\lambda_{\rm R50}$), the steepness and the top transmission efficiency for each band. In this paper, we set the slope of the intrinsic transmission curve to be constant, the steepness are defined as ${\rm Tan}_{x}=\Delta\lambda_x/\lambda_{x}$, here the subscript $x$ represents ${\rm L50}$ or ${\rm R50}$, $\Delta\lambda_x$ is the the wavelength difference between 0\% and 100\% maximum transmission. We estimate the limiting magnitude $m_{\rm Lim}$ for point sources with 5$\sigma$ detection at each band, here the point sources are measured within 80\% energy concentration region of a Gaussian PSF. Then we find that $m_{\rm Lim}^{\rm F275W}\simeq25.8$, $m_{\rm Lim}^{\rm F336W}\simeq26.2$, $m_{\rm Lim}^{\rm F375M}\simeq26.0$, $m_{\rm Lim}^{\rm F450M}\simeq26.4$, $m_{\rm Lim}^{\rm F500M}\simeq26.4$, $m_{\rm Lim}^{\rm F630M}\simeq25.9$, $m_{\rm Lim}^{\rm F763M}\simeq26.4$, $m_{\rm Lim}^{\rm F845M}\simeq26.2$ and $m_{\rm Lim}^{\rm F960M}\simeq24.9$ AB mag. This filter definition is the basic case, we will study the dependence of each MCI medium-band filter and the definition parameters on the photo-$z$ accuracy in our following discussion.

\begin{table*}
\centering
\caption{The CSST-MCI medium-band filter definition.}
\label{tab:filters}
\begin{tabular}{ c  c  c  c  c  c  c  c}
\hline\hline
Filter & $\lambda_{\rm c}$ $({\rm \AA})$ & FWHM $({\rm \AA})$ & $\lambda_{\rm L50}$ $({\rm \AA})$ & $\lambda_{\rm R50}$ $({\rm \AA})$ & ${\rm Tan}_{\rm L50}$ & ${\rm Tan}_{\rm R50}$ & Trans.\\
\hline
$F275W$ & 2725 & 450 & 2500 $\pm$ 25 & 2950 $\pm$ 25 & 0.05 & 0.02 & 65\%\\
$F336W$ & 3371 & 540 & 3101 $\pm$ 30 & 3641 $\pm$ 30 & 0.03 & 0.03 & 80\%\\
$F375M$ & 3750 & 500 & 3500 $\pm$ 35 & 4000 $\pm$ 35 & 0.02 & 0.02 & 80\%\\
$F450M$ & 4500 & 500 & 4250 $\pm$ 30 & 4750 $\pm$ 30 & 0.02 & 0.02 & 90\%\\
$F500M$ & 5000 & 500 & 4750 $\pm$ 30 & 5250 $\pm$ 30 & 0.02 & 0.02 & 90\%\\
$F630M$ & 6300 & 400 & 6100 $\pm$ 35 & 6500 $\pm$ 35 & 0.015 & 0.015 & 90\%\\
$F763M$ & 7647 & 790 & 7252 $\pm$ 40 & 8042 $\pm$ 40 & 0.015 & 0.015 & 90\%\\
$F845M$ & 8472 & 880 & 8032 $\pm$ 40 & 8912 $\pm$ 40 & 0.02 & 0.01 & 90\%\\
$F960M$ & 9600 & 600 & 9300 $\pm$ 40 & 9900 $\pm$ 40 & 0.01 & 0.01 & 90\%\\
\hline\hline
\end{tabular}
\end{table*}

\subsection{Galaxy catalog}
\label{sect:dat}

Following \cite{Cao18}, we use the COSMOS08 and COSMOS15 catalogs to generate the mock galaxy catalog for the CSST-MCI observation in the SC photo-$z$ calibration. The data of the catalogs cover about 2 deg$^2$ COSMOS field, which is expected to be similar as the CSST-MCI SC photo-$z$ calibration survey. The wavelength range is from the ultraviolet to mid-infrared (30 bands) from Galaxy Evolution Explorer (GALEX) \citep{Martin05}, Subaru Telescope \citep{Taniguchi07,Capak07},  Canada-France-Hawaii Telescope (CFHT) \citep{Boulade03}, and United Kingdom Infrared Telescope (UKIRT) \citep{Warren07} and Spitzer Space Telescope \citep{Rowan08}. The catalogs contain 385,065 sources with $i^+ \le 25.2$, and the photo-$z$ accuracy of $\sigma_{\Delta z}$ at $z<1.25$ is 0.02, 0.04, 0.07 for Subaru $i^+ \sim 24$, $\sim 25$, $\sim 25.5$, respectively.

In order to investigate the photo-$z$ calibration, we need to generate the mock flux data based on redshift, magnitude, galaxy type, best-fitting SED, dust extinction and other information from the catalogs. To collect all of these properties, we need to combine the COSMOS08 and COSMOS15 catalogs\footnote{The information of galaxy size is missing in the COSMOS08 catalog, while there are no dust extinction and best-fitting SED in the CSOMOS15 catalog. Hence we need to match the relevant galaxies in the two catalogs and combine them together to collect all of necessary information.}. Then 219,566 galaxies are obtained after removing stars, X-ray and masked sources. In order to perform the photo-$z$ estimation with required accuracy, we select the high-quality sources with the signal-to-noise ratio (SNR) $\ge$10 in $g$ or $i$ band \citep{Zhan21}. The calculation of SNR will be discussed in Section~\ref{sect:dat}. Then we obtain 103,931 galaxies as the selected catalog. Finally, we randomly select 10,000 galaxies from the high-quality sample, and calculate the photometric data and errors for each band based on the CSST instrumental parameters.

\begin{figure}
\centering
\includegraphics[width=1.\columnwidth]{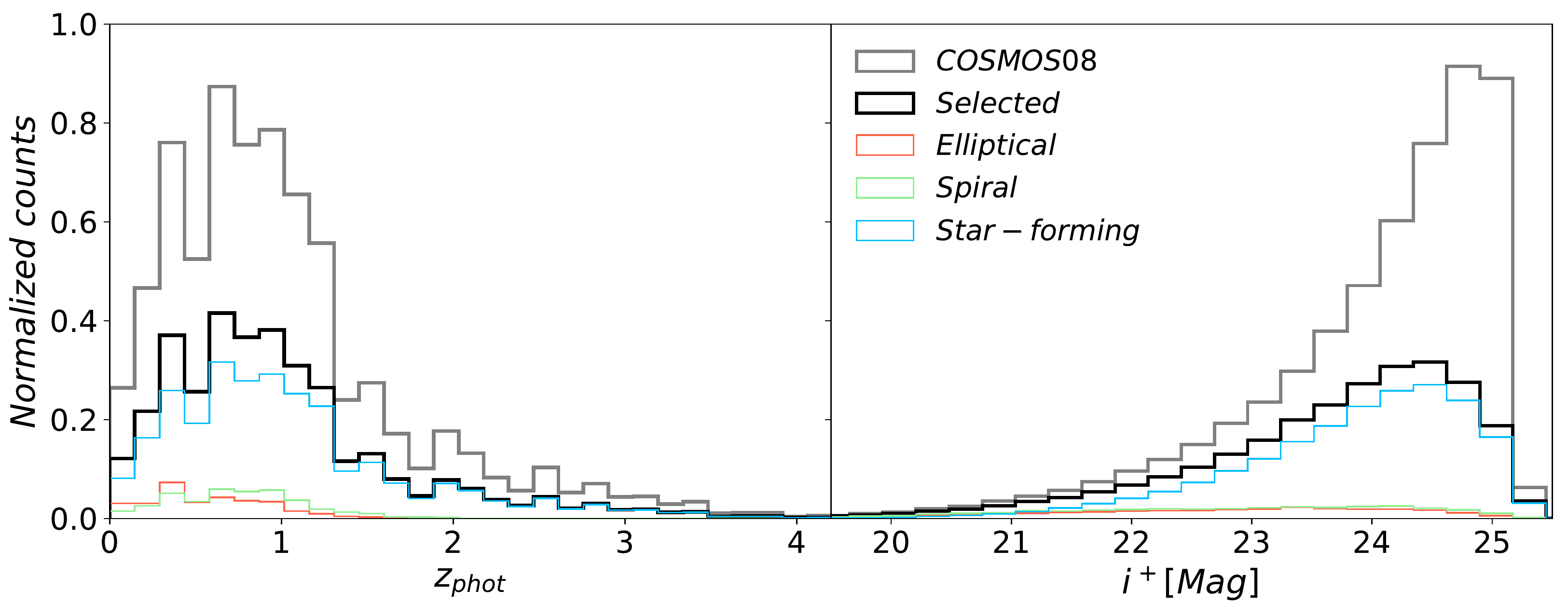}
\caption{The redshift and magnitude distributions of the galaxies in the mock catalog. The black and gray histograms show the galaxies from the selected high-quality samples and COSMOS08 catalog, respectively. The distributions of elliptical, spiral, and star forming galaxies are shown in red, green and blue histogram, respectively.
We can find that the peak of the redshift distribution is $z\sim0.7$ before and after the selection, and some faint galaxies are removed during the selection process. The peak of the magnitude distribution becomes $i\sim24.3$ after the selection. 
}
\label{fig:cat}
\end{figure}

In Figure~\ref{fig:cat}, we show the magnitude and redshift distributions of the galaxies we use in the left and right panels. Here, the black and gray histograms show the galaxy samples from the selected catalog and the COSMOS08 catalog, respectively. We can find that some faint galaxies are removed during the selection process, so the peak of the magnitude distribution becomes $i\sim24.3$ after selecting, but the range is still from $i\sim19$ to $\sim25$. The redshift distribution of the two catalogs is similar, and the peak of redshift distribution around $z=0.7$ and the range can extend to $z\sim4$. In COSMOS08 catalog, galaxies are classified into three categories: elliptical, spiral, and young blue star-forming galaxies, shown in red, green and blue histograms, respectively. We find that the star forming galaxies are dominant in the catalog, accounting for about 75$\%$ of the total. For the star forming and spiral galaxies, the peaks of redshift distributions are at $z\sim0.7$, and it is at $z\sim0.3$ for elliptical galaxies.

\subsection{Data and error estimation}
\label{sect:dat}

We adopt the SED templates used in the COSMOS photometric redshift fitting as the original SED templates to generate the mock flux data \citep{Arnouts99,Ilbert06}. Totally 31 basic SED templates are used here, including seven elliptical galaxy templates and twelve spiral galaxy templates (from S0 to Sdm), which are derived from \cite{Polletta07}. Other templates are the young star-forming galaxy templates (starburst ages from 0.03 to 3 Gyr), which are generated by the BC03 method \citep{Ilbert09}. Following \cite{Cao18}, we extend the wavelength coverage of the COSMOS templates to $\sim90$ $\rm\AA$ using the BC03 models. 
In order to increase the completeness of the SED templates and avoid the ''over-fitting'' effect, we use the linear interpolation method to generate new SEDs. 
For simplicity, we assume that the transitions between different SEDs are continuous and that the probability of adjacent SEDs is similar, so that the the new intrinsic galaxy SED $S_{\rm int}$ can be written as,
\begin{equation}
S_{\rm int} = (1-a) S_{\rm i} + a\,S_{\rm i\pm1},
\end{equation}
where, $a \in (-0.5,0.5)$ is a random factor, $S_{\rm i}$ is the best-fit SED template for a source from the COSMOS catalog, and $S_{\rm i\pm1}$ is the adjacent templates of $S_{\rm i}$. We set $S_{\rm i+1}$ for $a>0$, and $S_{\rm i-1}$ when $a<0$.


An important effect to consider is the extinction of dust from the galaxies themselves, especially for high redshift galaxies. The most used extinction law in high redshift studies is derived by \cite{Calzetti00} for the near starburst galaxy (SBG). To increase the diversity, four other excellent dust extinction laws are also adopted, they are derived from the studies of the Milky Way (MW) \citep{Allen76,Seaton79}, Large Magellanic Cloud (LMC) \citep{Fitzpatrick86}, Small Magellanic Cloud (SMC) \citep{Prevot84,Bouchet85}. We add the dust extinction to SED when generating the mock flux data. Here the flux density after dust reddening form galaxies themselves can be expressed as \citep{Calzetti94,Galametz17},
\be \label{eq:SED_ext}
S_{\rm ext}(\lambda_{\rm res}) = S_{\rm int}(\lambda_{\rm res})\,10^{-0.4E(B-V)k(\lambda_{\rm res})},
\ee
Here $E(B-V)$ is the color excess, $k(\lambda)$ is the reddening curve. Another effect that must be accounted for is the absorption of intergalactic medium (IGM). The emission from high redshift galaxies usually can be absorbed by the neutral hydrogen clouds in the IGM. We make use of the attenuation model discussed by \citet{Madau95} to process the redshifted $S_{\rm ext}$, and obtain the final observed spectrum $S_{\rm model}$. The extinction laws used above can be found in Figure.4 of \cite{Cao18}.

For a given spectrum from the target source, the value of mock flux can be calculated by convolving the final observed spectrum $S_{\rm model}$ with the filter transmission function $T(\lambda)$, so the mock flux $F^{\rm mock}$ can be defined as,
\begin{equation}
\label{eq:f_obs}
F^{\rm mock}_x = \int S_{\rm model}(\lambda) T_x(\lambda)\tau_x d\lambda.
\end{equation}
Here $x$ represents the $x$ band, $\lambda=\lambda_{\rm res}(1+z)$ is the observed wavelength, and $\tau$ is the mirror efficiency. We show the CSST mirror efficiency in Table~\ref{teb:para}. For the CSST-MCI, the mirror efficiency is found to be $\sim$0.5 for $F275W$ and $F336W$ band, and $\sim$0.6 for other bands. Since $S_{\rm model}$ is a normalized spectrum, we need to use the information of the COSMOS galaxy catalog to calibrate the mock data. We can write the AB magnitude $m_x$ on the $x$ band as,
\begin{equation}
m_x = -2.5\log_{10}\left(\frac{F^{\rm mock}_{x}}{F^{\rm mock}_{i^+}}\right)+m_{i^+}
\end{equation}
Where $m_{i^+}$ is the magnitude of the Subaru $i^+$ band in the COSMOS galaxy catalog.

\begin{table*}
\centering
\caption{The parameters of the CSST MCI and SC filters.}
\label{teb:para}
\begin{tabular}{ c  c  c  c  c  c  c  c c c}
\hline\hline
Filter& $F275W$ & $F336W$ & $F375M$ & $F450M$ & $F500M$ & $F630M$ & $F763M$ & $F845M$ & $F960M$\\
$m_{\rm Lim}$& 25.8 & 26.2 & 26.0 & 26.4 & 26.4 & 25.9 & 26.4 & 26.2 & 24.9\\
$N^{\rm sky}$& 0.001 & 0.006 & 0.009 & 0.035 & 0.043 & 0.042 & 0.079 & 0.074 & 0.020\\
$\tau$&0.5&0.5&0.6&0.6&0.6&0.6&0.6&0.6&0.6\\
\hline
Filter& $NUV$ & $u$ & $g$ & $r$ & $i$ & $z$ & $y$\\
$m_{\rm Lim}$& 25.4 & 25.4 & 26.3 & 26.0 & 25.9 & 25.3 & 24.5 \\
$N^{\rm sky}$& 0.003 & 0.017 & 0.160 & 0.205 & 0.212 & 0.127 & 0.037\\
$\tau$&0.55&0.67&0.82&0.82&0.82&0.82&0.82\\
\hline\hline
\end{tabular}
\end{table*}

We now estimate the photometric error measured for the CSST. Here we use an approximate relation $\sigma_{\rm ph}\simeq 2.5\,{\rm log_{10}}\left[ 1+1/{\rm SNR}\right]$ to evaluate the magnitude error \citep{Bolzonella00,Pozzetti96,Pozzetti98}, then we add a systematic error $\sigma_{\rm sys}=0.02$ mag for the mock data, and the total photometric error can be defined by $\sigma_m=\sqrt{\sigma_{\rm ph}^2+\sigma_{\rm sys}^2}$. According to \cite{Cao21}, we find that the noise of each band satisfies the Gaussian distribution, so we add a Gaussian error ${\rm N}(0,\sigma_m^2)$ to the magnitude of the mock data.

For a space telescope, after $n$ exposures with time $t$, the signal to noise ratio (SNR) at wavelength $\lambda$ can be evaluated by \citep{Ubeda12,Cao18},
\begin{equation} 
\label{eq:SNR}
{\rm SNR} = \frac{\sqrt{n}C_{\lambda}\,t}{\sqrt{C_{\lambda}\,t+n_{\rm pix}(N_{\lambda}^{\rm sky}\,t+N^{\rm dark}\,t+N_{\rm RN}^2})},
\end{equation}
here, the number of exposures $n$ are 8 and 2 for the MCI and SC, respectively, the single exposure time $t=300{\,\rm s}$ for MCI and 150 s for SC, $n_{\rm pix}$ is the number of detector pixels used in spectral sampling, $N_{dark}=0.017\ e^-{\rm /s}/{\rm pixel}$ and $N_{\rm RN}=5.5\ e^-/{\rm pixel}$ are the dark current and the read noise of CCD, respectively.  $N_{\lambda}^{\rm sky}$ is the sky background at at wavelength $\lambda$, we estimate the sky background $N^{\rm sky}$ based on the average values of the zodiacal light and earthshine measured by \citet{Ubeda12}, and show of $N^{\rm sky}$ the values for each filter in Table~\ref{teb:para}.

In Equation~\ref{eq:SNR}, $C_{\lambda}$ is the count rate from the target source at wavelength $\lambda$ in $e^-/\rm s$, and it can be expressed as,
\begin{equation}
C_{\lambda}=10^{-0.4(m-{\rm ZP})}
\end{equation}
where, $m$ is the magnitude of an arbitrary object. ${\rm ZP}$ is the zero-point of an instrument, and we can define the zero-point of a filter such that \citep{Cao21},
\begin{equation}
{\rm ZP}=m_{\rm Lim}+2.5\log \left({\rm SNR}\sqrt{n_{\rm pix}}\sigma\right)
\end{equation}
where, $m_{\rm Lim}$ is the limiting magnitude, and we show the values of $m_{\rm Lim}$ for each filter in Table~\ref{teb:para}. $\sigma$ is the standard deviation of noise. 

\begin{figure*}
\centering
\includegraphics[width=1.\columnwidth]{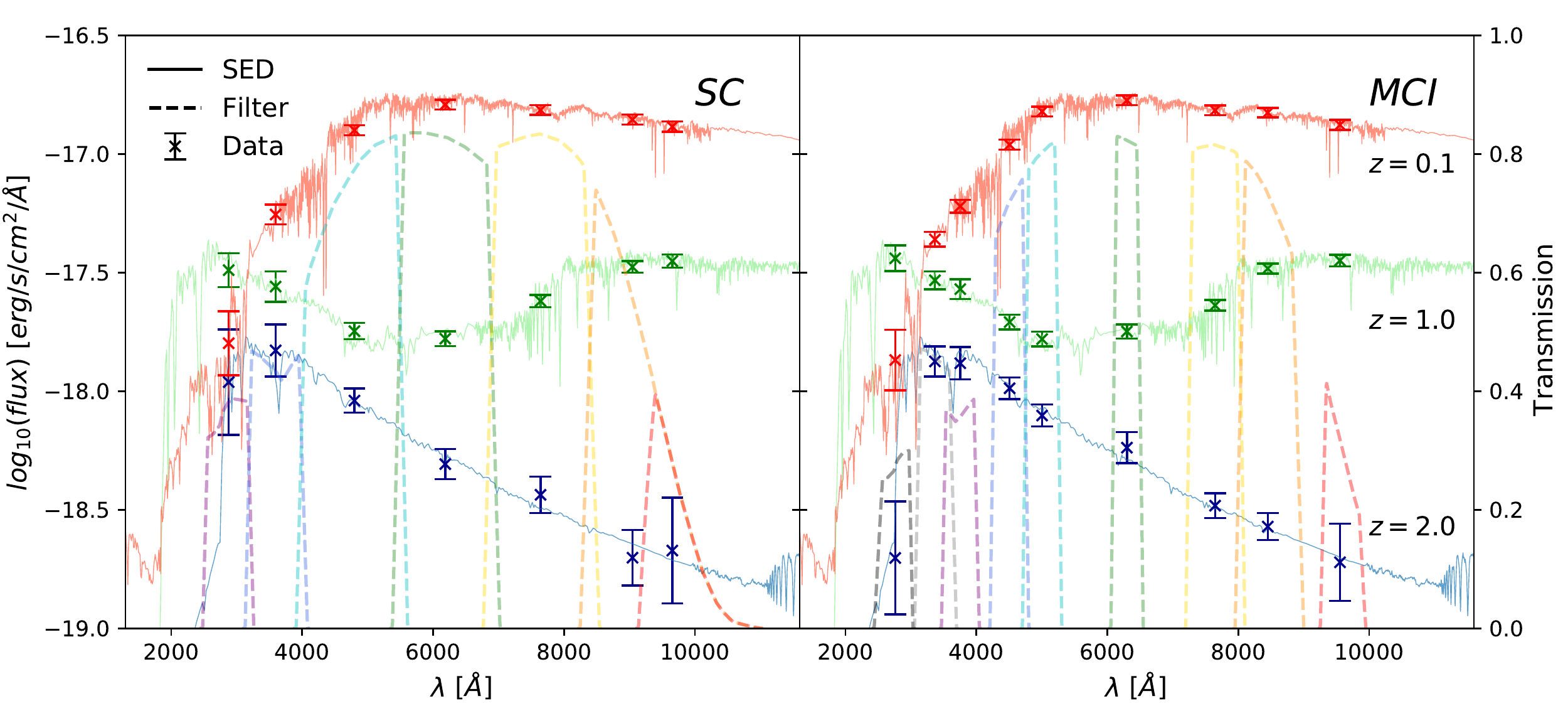}
\caption{\label{fig:flux} An example of mock flux data of four randomly selected galaxies with redshift $z\simeq0.1$, $1.0$ and $2.0$. Here, the dashed lines show the filter transmission with detector quantum efficiency, the solid lines denote the final observed SED, and the crosses with error bars are the mock flux data. The SC and MCI data are shown in the left and right panels, respectively.}
\end{figure*}

In Figure \ref{fig:flux}, we show an example of mock flux data from four randomly selected galaxies with redshifts $z\simeq0.1$, $1.0$ and $2.0$. The SC and MCI data are shown in the left and right panels, respectively. Here, the crosses and solid lines are the mock flux data and final observed spectra described in Equation~\ref{eq:f_obs}, respectively. The error bars are 1$\sigma$ uncertainties that account for all noises and errors.

\section{Photo-$z$ Estimation and filter dependency}
\label{sect:est}

\subsection{Photo-$z$ accuracy estimation}

Following \cite{Cao18}, we modify the LePhare code to include the upper-limit information \citep{Arnouts99,Ilbert06}. Here the best fitting redshift $z_{\rm bf}$ is derived by the chi-square $\chi^2$ method, which can be obtained by, 
\begin{equation}
\label{eq:chi_N}
\chi^2 = \sum_{i}^{N}\left( \frac{m_i^{\rm obs}-m_i^{\rm pre}}{\sigma_i^{\rm obs}} \right)^2 +  \sum_j^M -2\log(P_j),
\end{equation}
where $N$ and $M$ are the number of data with ${\rm SNR}\ge3$ and ${\rm SNR}<3$, respectively. $m_i^{\rm obs}$ and $\sigma_i^{\rm obs}$ are the observed AB magnitude and its error for the $i$-th band, respectively, which are derived from the mock data described in the Section~\ref{sect:dat}. $m_i^{\rm pre}$ is the predicted magnitude by the photo-$z$ fitting software. $P_j$ is the probability that the predicted data is between 0 and the upper limit value proposed by \cite{Cao18},
\begin{equation}
P_{j}=\frac{1}{\sqrt{2 \pi} \sigma_{j}} \int_{0}^{F_{\mathrm{u}}} \exp \left[-\frac{\left(f-F_{j}^{\mathrm{th}}\right)^{2}}{2 \sigma_{j}^{2}}\right] df.
\end{equation}
Here $f$ is the flux variable, $F_{j}^{th}$ is the predicted flux in band $j$, $\sigma_{j}$ is the flux error, and $F_{\mathrm{u}}=3\sigma_{j}$ is the upper limit value of band $j$.

Here, we use two parameters: the deviation $\sigma_{z}$ and the catastrophic redshift fraction $f_c$ to analyze the accuracy of redshift. The catastrophic redshift fraction $f_c$ represents the proportion of data with $|\Delta z|/(1+z_{\rm in})>0.15$ to the total number, where $\Delta z$ is the difference between the input redshift $z_{\rm in}$ and the photo-$z$ best fitting redshift $z_{\rm bf}$. In order to suppress the weighting of a small number of large error data, we use the normalized median absolute deviation $\sigma_{\rm NMAD}$ \citep{Ilbert06,Brammer08} to calculate the deviation $\sigma_{z}$ of $\Delta z$, and we can write as,
\begin{equation}
\sigma_{z}=\sigma_{\rm NMAD} = 1.48 \times {\rm median}\left( \left| \frac{\Delta z-{\rm median}(\Delta z)}{1+z_{\rm input}}\right|\right).
\end{equation}

\begin{figure*}
\centering
\includegraphics[width=0.33\columnwidth]{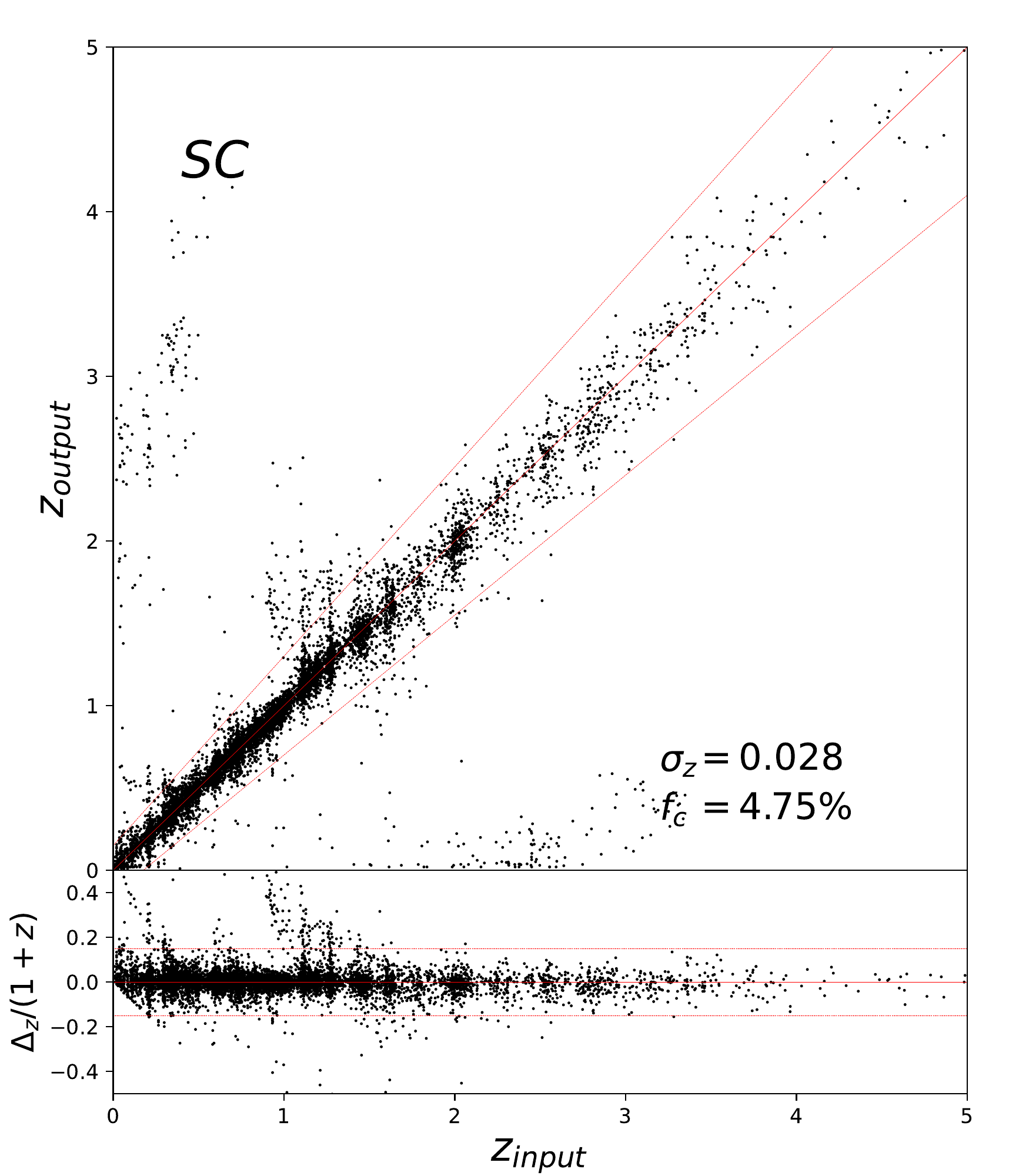}
\includegraphics[width=0.33\columnwidth]{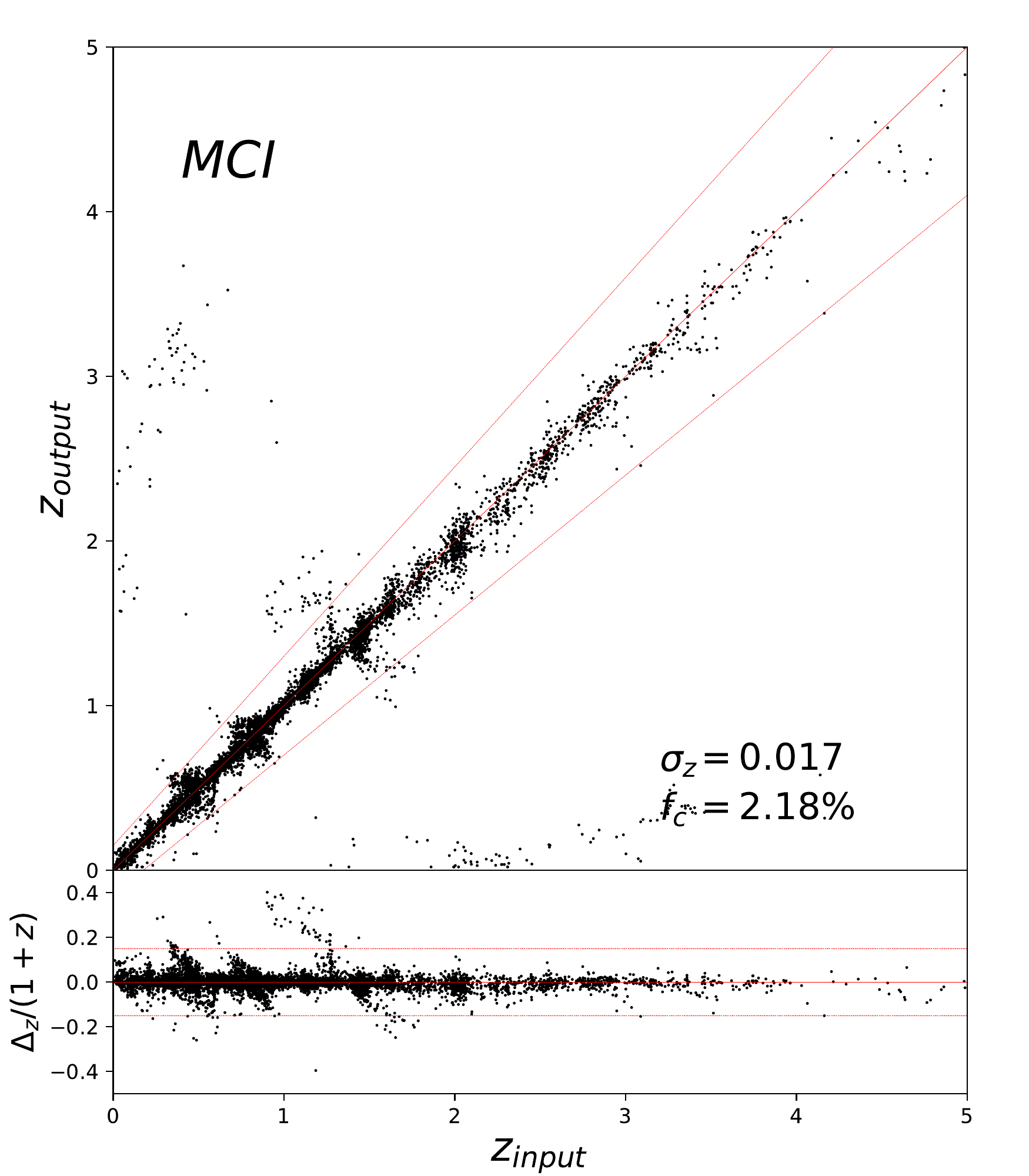}
\includegraphics[width=0.33\columnwidth]{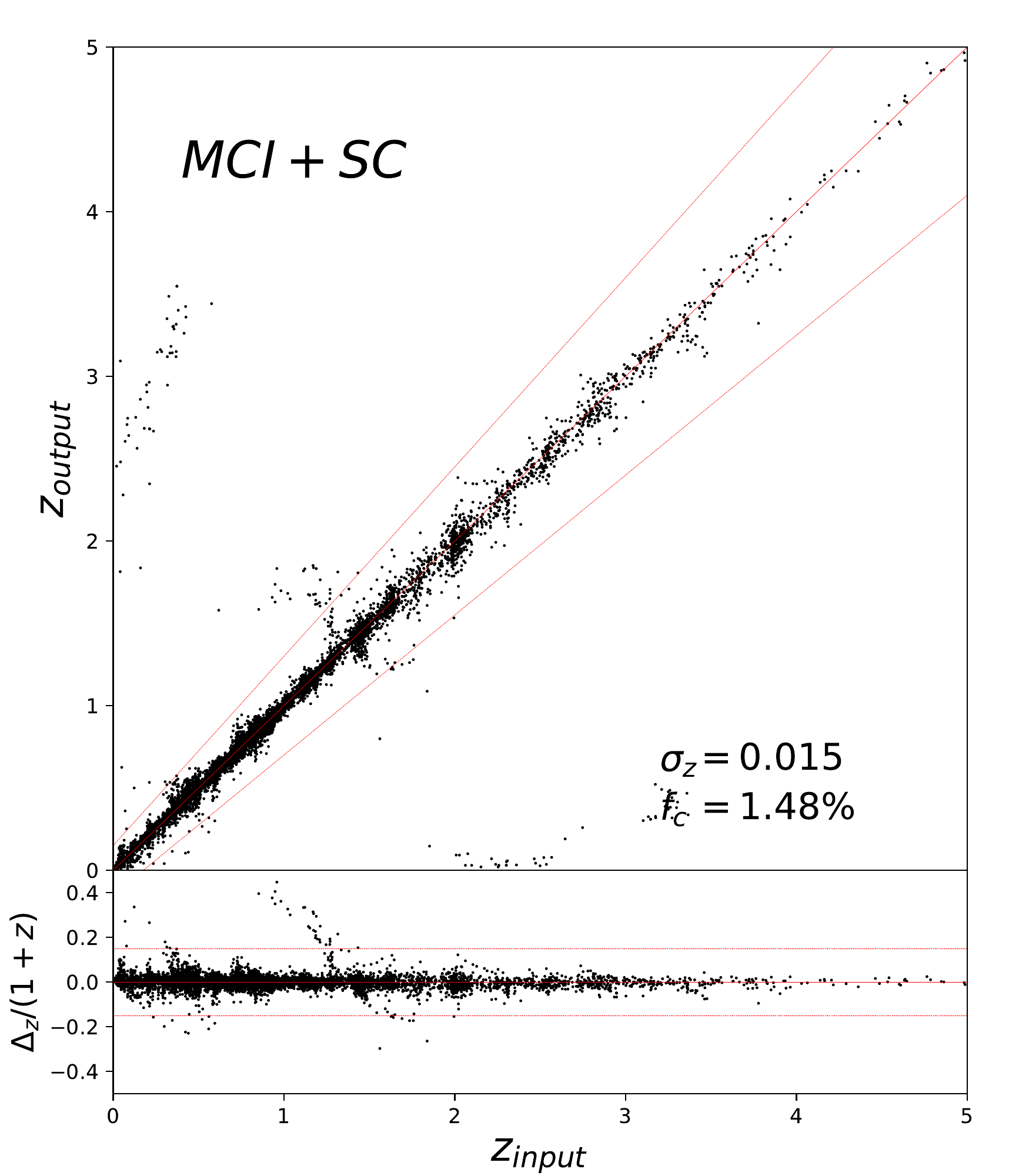}
\caption{\label{fig:all} $Left:$ The photo-$z$ fitting results including the information of upper-limits. Here, the photo-$z$ are obtained from the SC data only. 
$Middle:$ The result obtained from the MCI data only. There are a few areas of poor estimation due to gaps between bands, which will be discussed in detail in Section~\ref{sect:band}.
$Right:$ The result including all SC and MCI data. Here the deviation $\sigma_z=0.015$ and the catastrophic data redshift fraction $f_{\rm c}=1.48\%$, which is significantly improved compared to $\sigma_z=0.028$ and $f_{\rm c}=4.75\%$ without considering the MCI data.}
\end{figure*}

We first investigate the improvement of photo-$z$ accuracy by adding MCI medium-band filters. The photometric data are obtained by convolving all the MCI medium-band filters and SC filters with SED based on the method introduced in Section~\ref{sect:dat}, and then we estimate and analyze the photo-$z$ of the mock photometric data with and without MCI data. We show $z_{\rm input}$ vs. $z_{\rm output}$ for the current filters in Figure~{\ref{fig:all}}, the left panel is the result with SC data only, and the right panel is the result with all data. We find that, the deviation and the catastrophic redshift fraction of the fitting results containing only the SC data are $\sigma_z=0.028$ and $f_{\rm c}=4.75\%$. After adding MCI data, we find that the accuracy is greatly improved compared with before, and the deviation and catastrophic redshift fraction become $\sigma_z=0.015$ and $f_{\rm c}=1.48\%$. 

Note that active galactic nucleus (AGNs) may contaminate galaxies in photometric surveys, that can result in wrong photo-$z$ estimation if they are misidentified as galaxies. However, we find that the number of AGNs ($\sim10^7$) observed by the CSST only takes about 1\% of the number of galaxies ($\sim10^9$) in the CSST photometric survey \citep{LSST09,Gong19}. Besides, since the CSST can simultaneously perform photometric and spectroscopic surveys, it has a distinct advantage of identifying and confirming AGNs, which can be helpful to effectively remove AGNs from the galaxy sample. Therefore, we assume that the AGN contamination can be ignored here.

\subsection{Dependency of photo-$z$ accuracy on each band}
\label{sect:band}

\begin{figure}
\centering
\includegraphics[width=0.31\columnwidth]{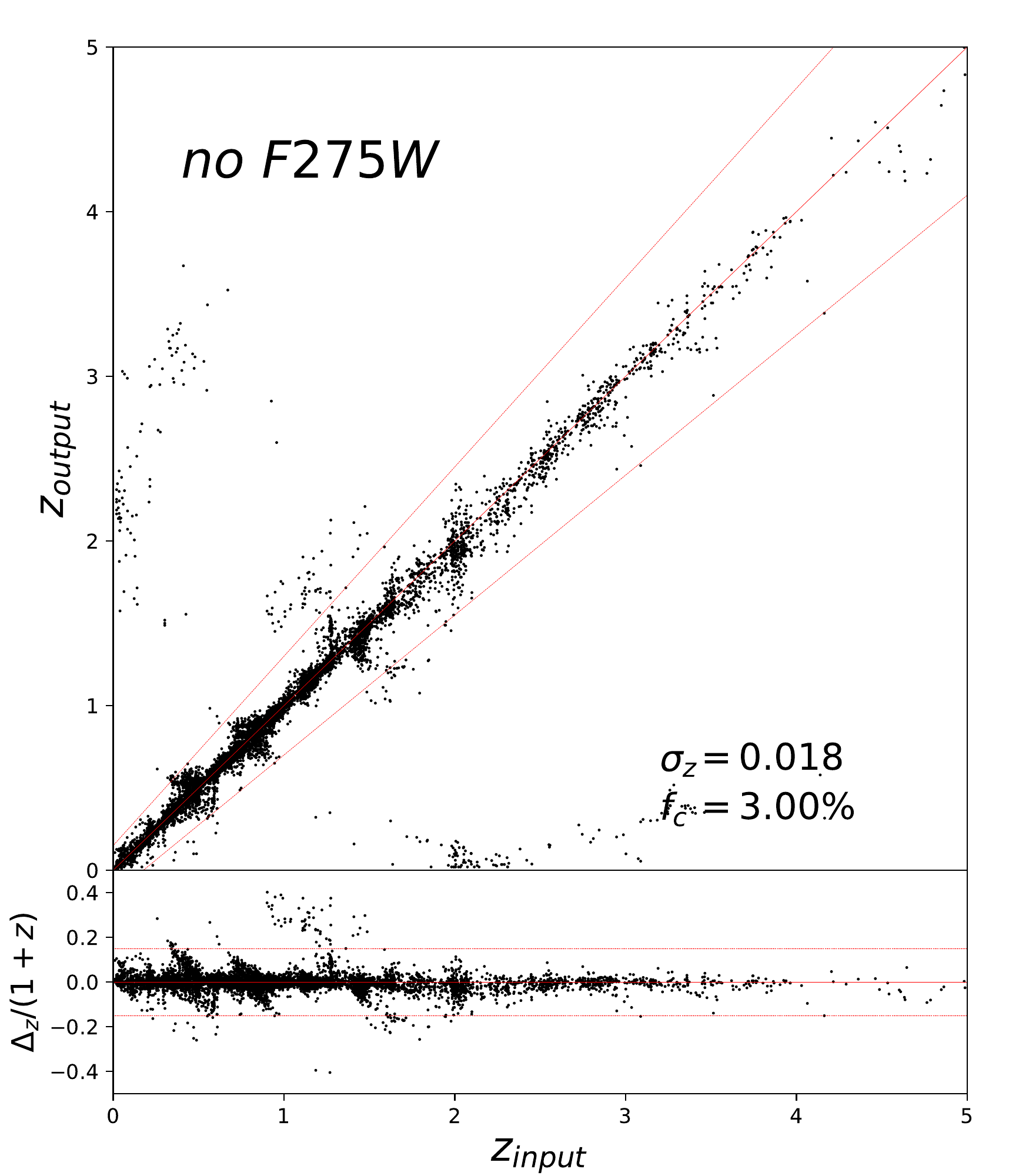}
\includegraphics[width=0.31\columnwidth]{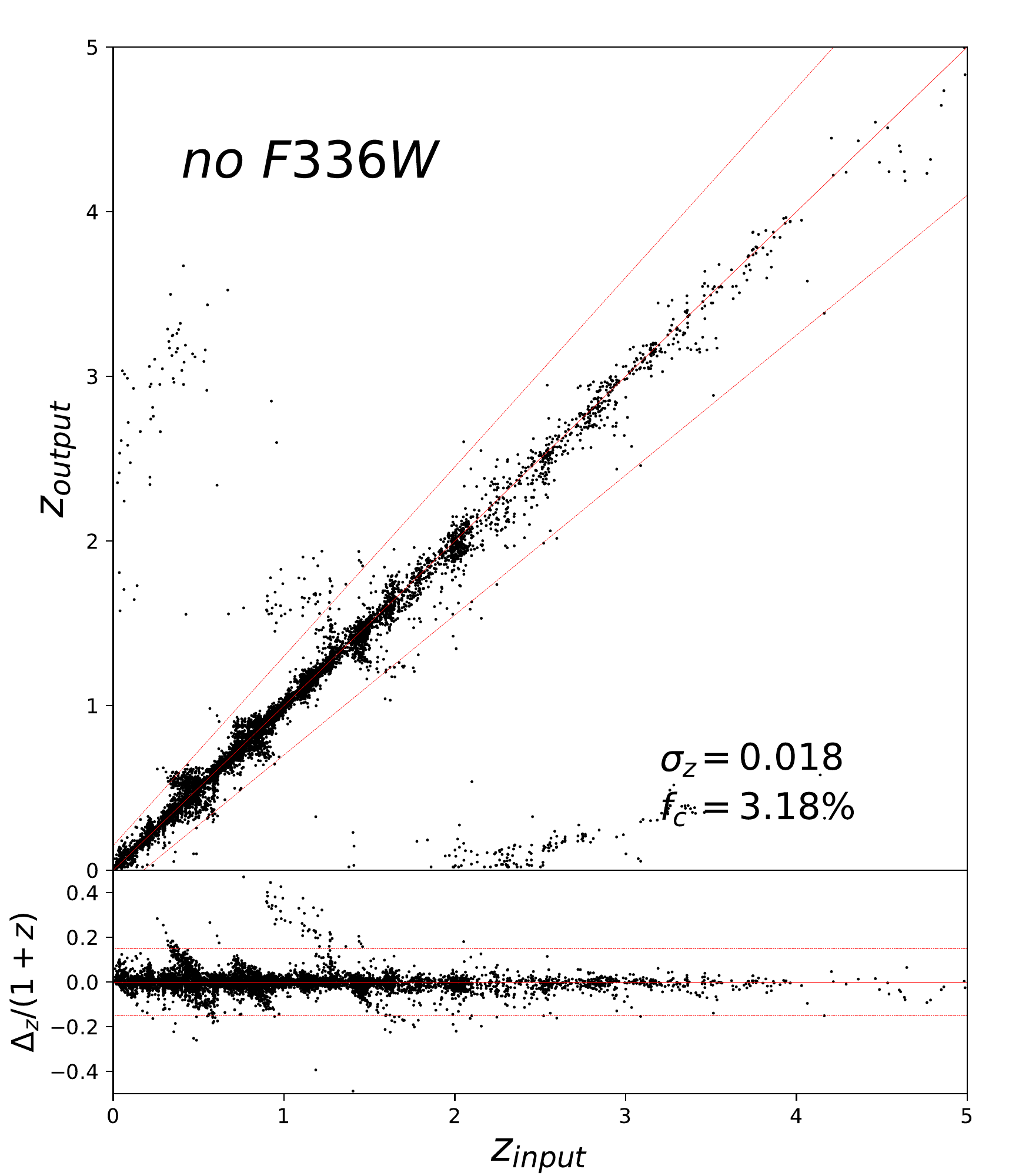}
\includegraphics[width=0.31\columnwidth]{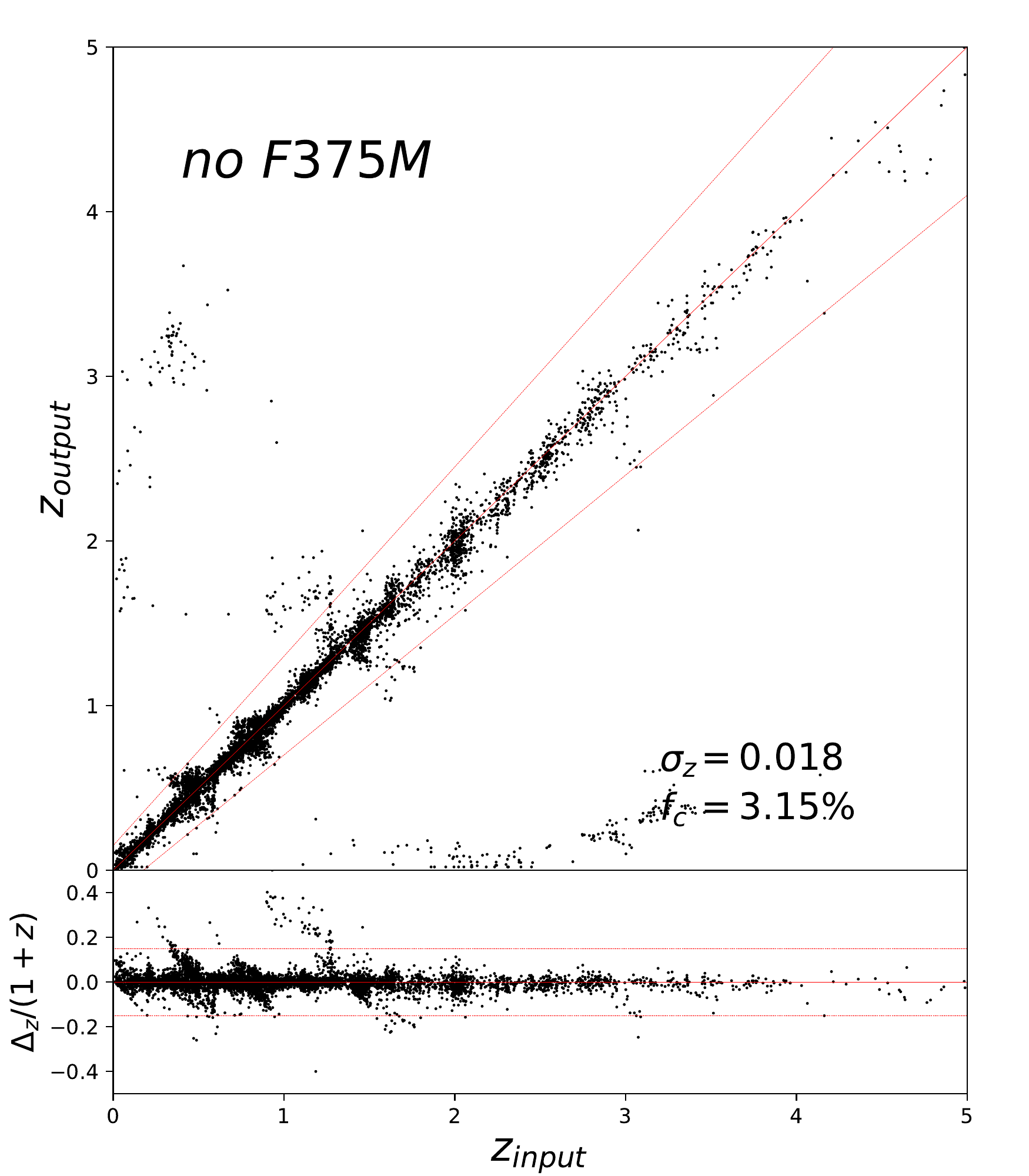}
\includegraphics[width=0.31\columnwidth]{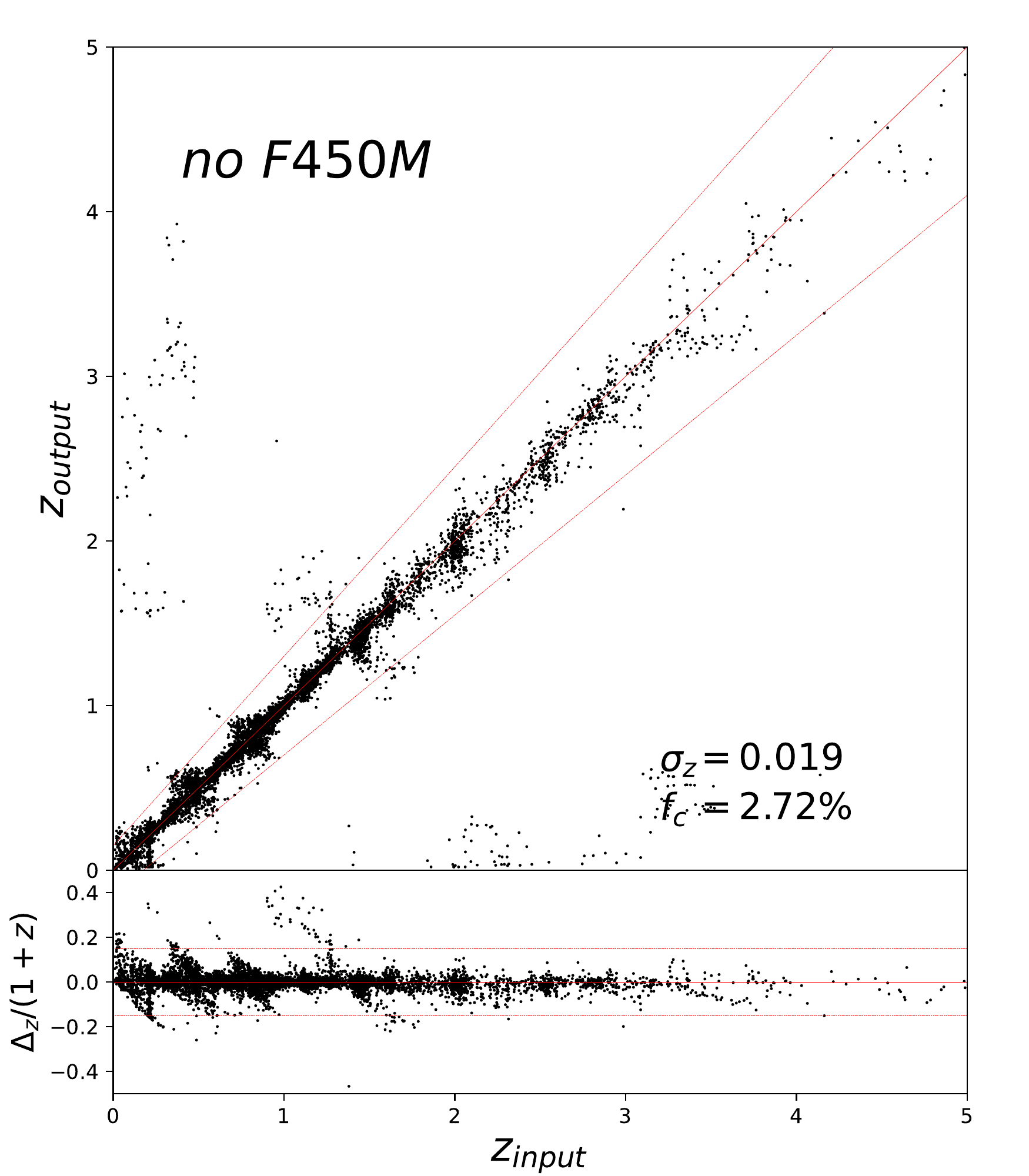}
\includegraphics[width=0.31\columnwidth]{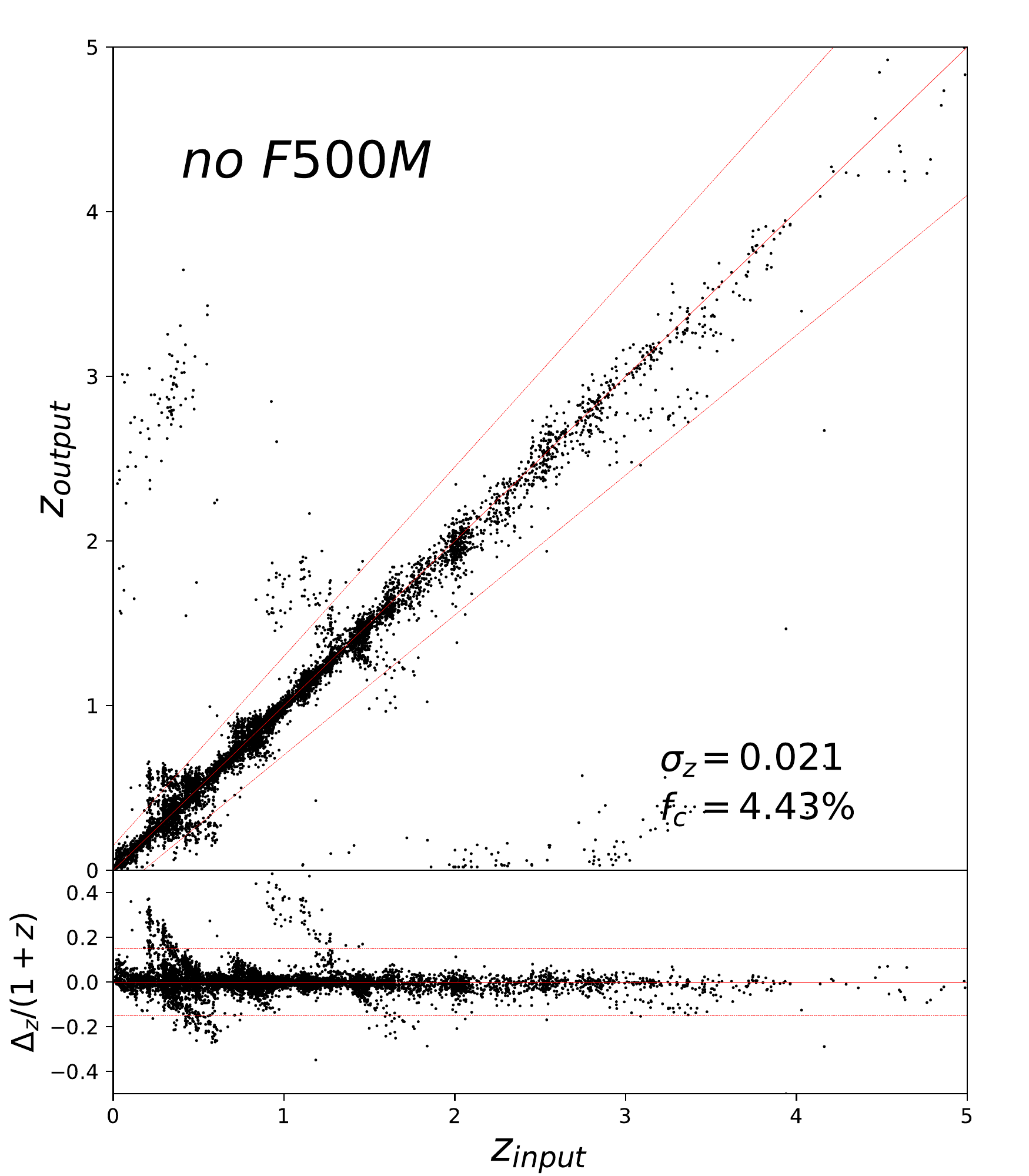}
\includegraphics[width=0.31\columnwidth]{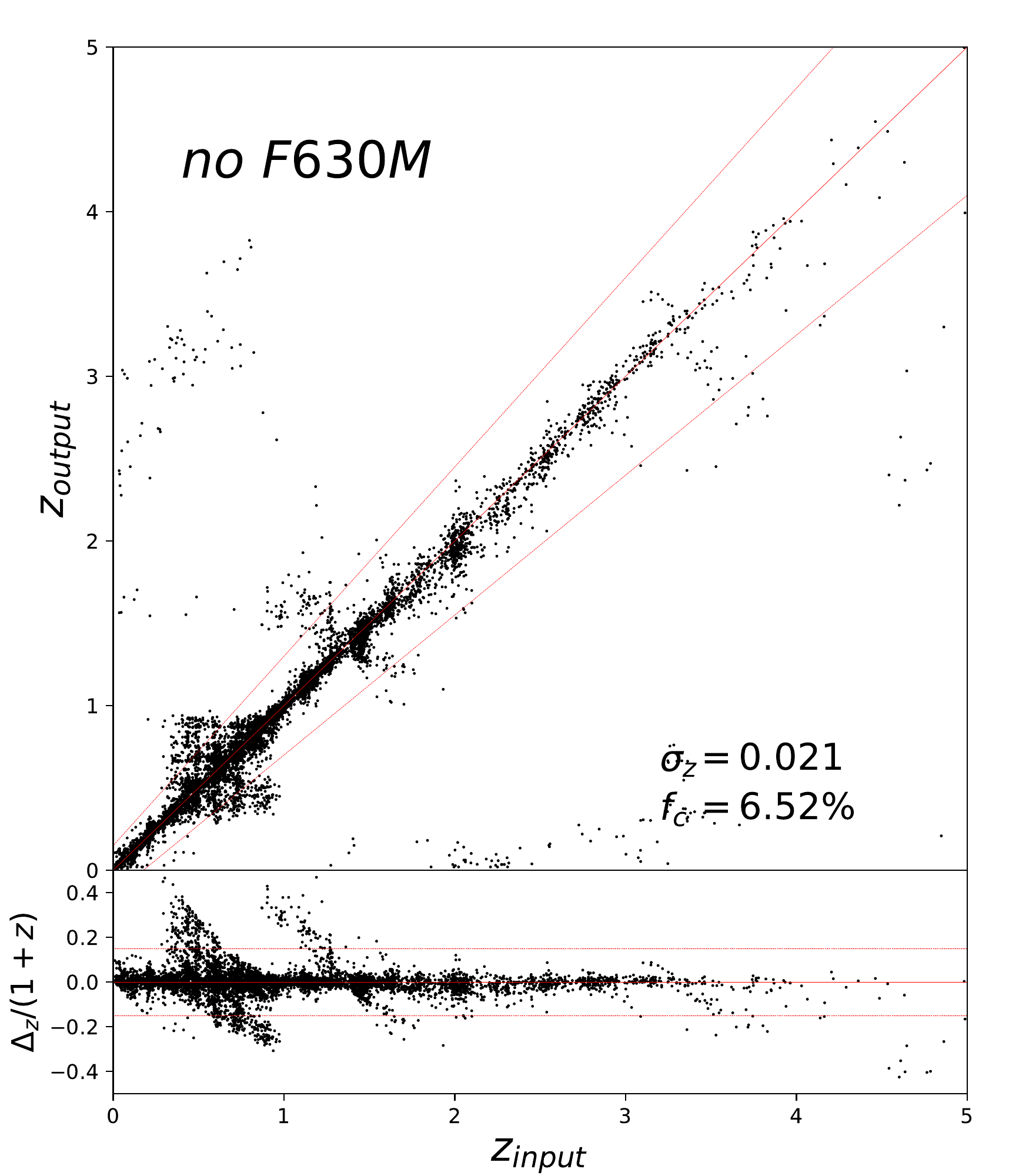}
\includegraphics[width=0.31\columnwidth]{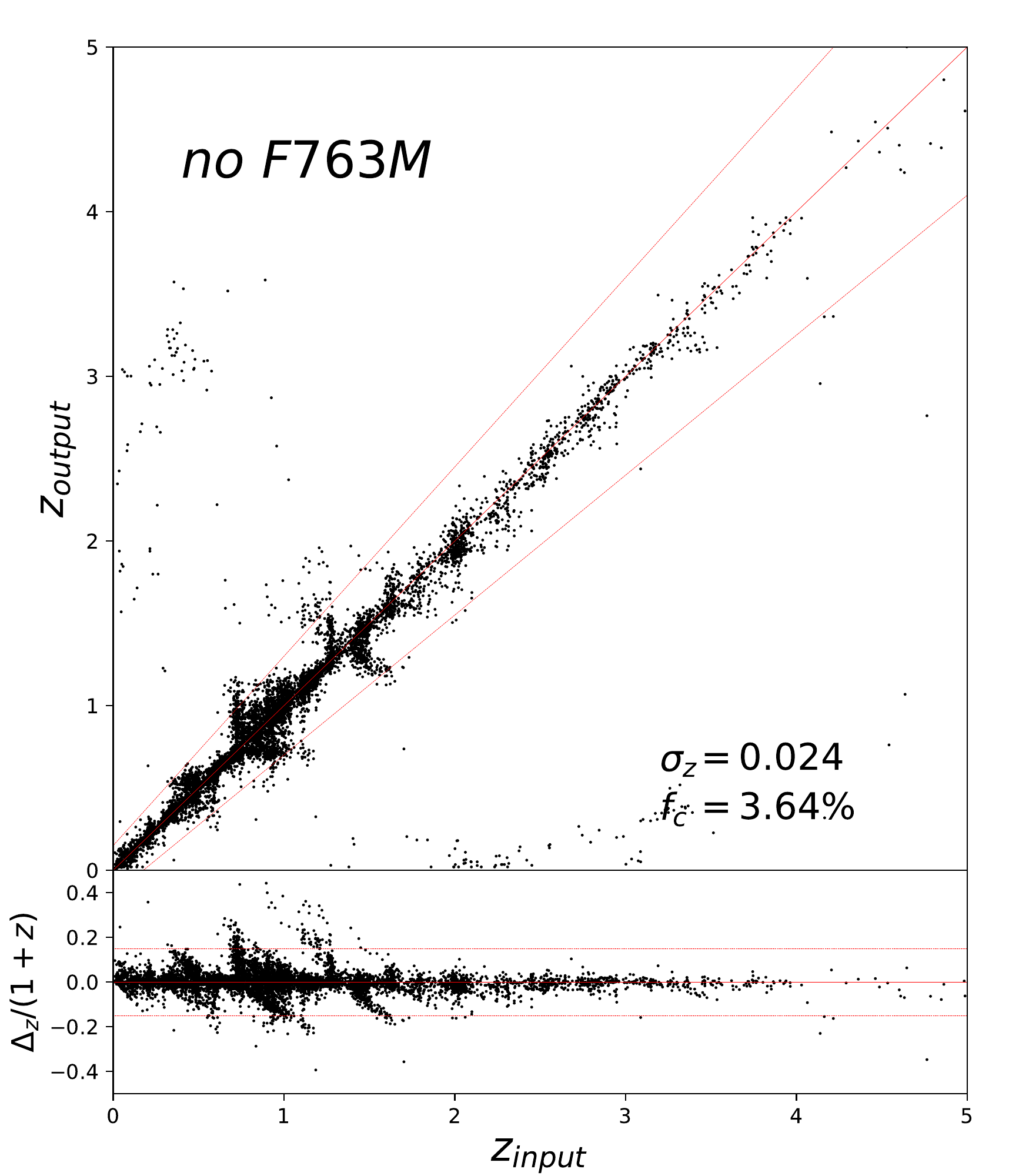}
\includegraphics[width=0.31\columnwidth]{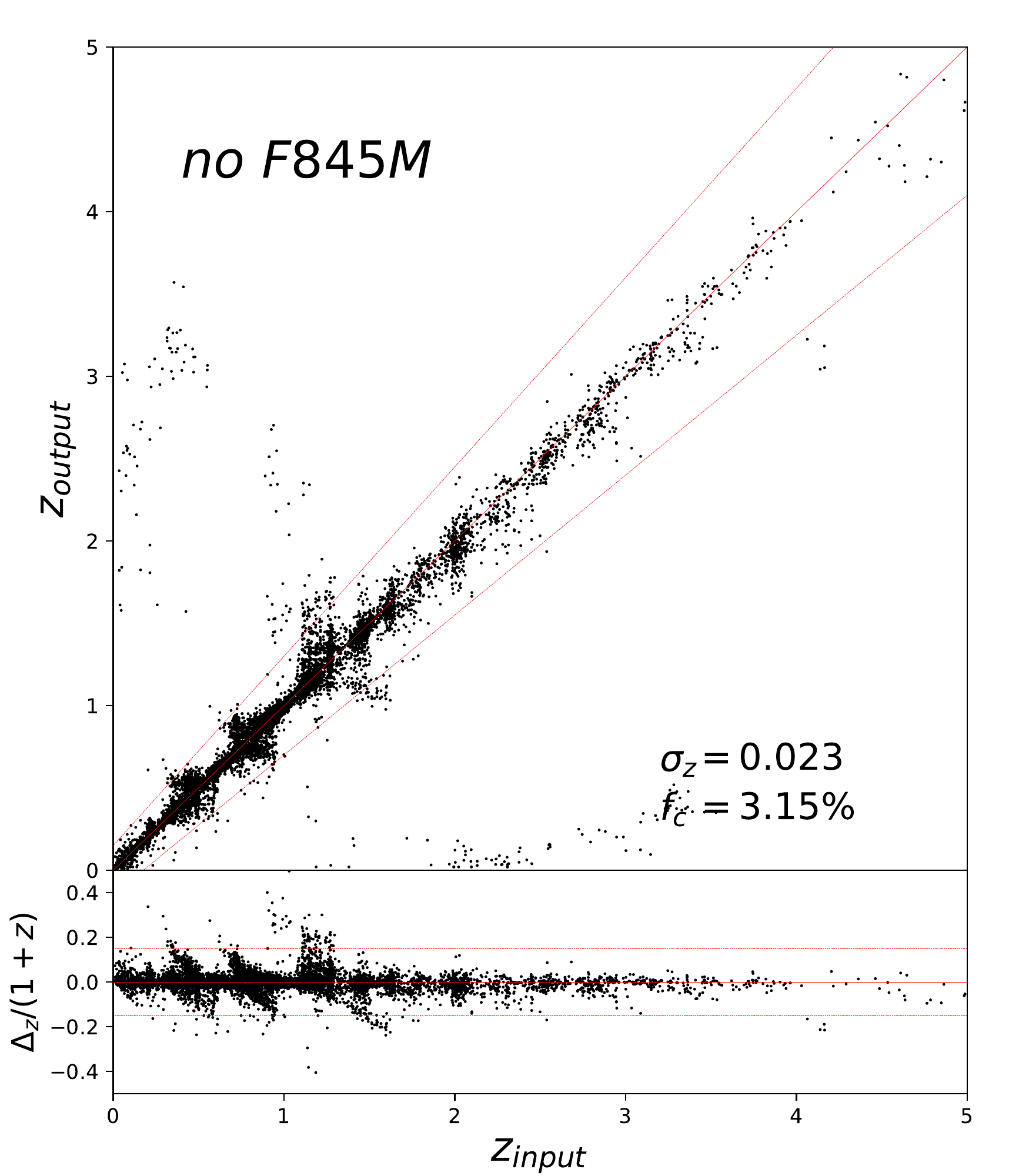}
\includegraphics[width=0.31\columnwidth]{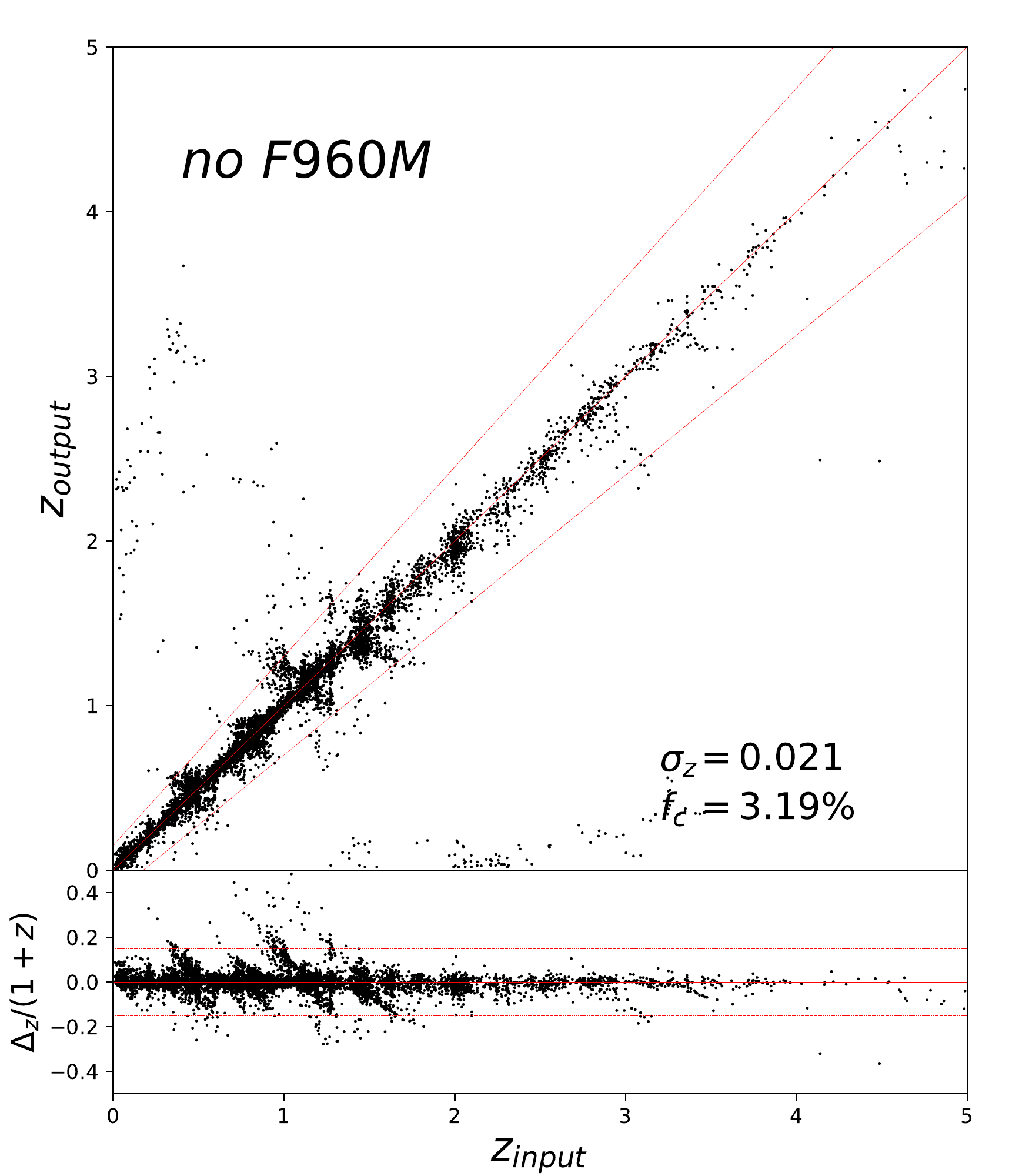}
\caption{The $z_{\rm input}$ vs. $z_{\rm output}$ for exploring the effect of a single MCI filter on the photo-$z$ accuracy. Due to the restriction of the efficiency of the filters in the near-ultraviolet band, they have less effect on the photo-$z$ accuracy than other bands. Removing the band with wavelengths greater than 4000\AA\ will result in a poor estimation in the corresponding redshift range.}
\label{fig:band} 
\end{figure}

Next we investigate the effect of each MCI medium-band filter passband on photo-$z$ accuracy by removing it, and explore the trend of the photo-$z$ accuracy with the change of filter parameters.
The value of the deviation $\sigma_z$ and the catastrophic redshift fraction $f_{\rm c}$ for removing one filter case are shown in the Figure~\ref{fig:band} and Table~\ref{tab:band}. Note that we divide the  results into two groups, $\sigma_z^*$ and $f_{\rm c}^*$ represents the result using only the MCI data, and $\sigma_z$, $f_{\rm c}$ represents the result using both the MCI and SC data. 

When removing the $F275W$, $F336W$ or $F375M$ band, we find that the values of deviation $\sigma_z^*=0.018$ (or $\sigma_z=0.015$) are similar to in the case that the relevant filter is included, but the catastrophic redshift fraction $f_{\rm c}^*$ (or $f_{\rm c}$) increases significantly. Due to the low efficiency of the filter in the near-ultraviolet band, the improvement of photo-$z$ accuracy is limited compared with other bands. However, the presence of the near-ultraviolet band can pin down the number of catastrophic redshifts \citep{Cao18}. This is because the accuracy of the template fitting method depends on the recognition and calibration of the strong spectral features. When fitting the values of photo-$z$, the fitting program may misidentify the Lyman break with the Balmer break or 4000\AA\ break. Adding detection information in the near-ultraviolet band helps improving the accuracy of recognition. On the other hand, improving the efficiency and increasing the exposure time can also improve the accuracy of photo-$z$ measurement.

\begin{table*}
\centering
\caption{The $\sigma_z$ and $f_{\rm c}$ for different filter sets.}
\label{tab:band}
\begin{tabular}{ c  c  c  c  c  c  c  c c c c}
\hline\hline
\multirow{2}{*}{Filters} & \multirow{2}{*}{all}&\multicolumn{9}{c}{Lack of the following filters, respectively.}\\
&& F275W & F336W & F375M & F450M & F500M & F630M & F763M & F845M & F960M\\
\hline
$\sigma_z^*$& 0.017 & 0.018 & 0.018 & 0.018 & 0.019 & 0.021 & 0.021 & 0.024 & 0.023 & 0.021\\
$f_{\rm c}^*$& 2.18 & 3.00 & 3.18 & 3.15 & 2.72 & 4.43 & 6.52 & 3.64 & 3.15 & 3.19 \\
\hline
$\sigma_z$& 0.015 & 0.015 & 0.015 & 0.015 & 0.015 & 0.016 & 0.016 & 0.017 & 0.017 & 0.015 \\
$f_{\rm c}$& 1.48 & 1.62 & 2.05 & 1.86 & 1.52 & 1.77 & 1.58 & 1.54 & 1.53 & 1.51\\
\hline\hline
 \multicolumn{11}{l}{The superscript $*$ indicates that the results are derived from the mock data without   the SC filters.}
 \end{tabular}
\end{table*}

As shown in Table~\ref{tab:band} and Figure~\ref{fig:band}, the bands with wavelengths greater than 3750\AA\ can significantly affect the photo-$z$ fitting results. The SED template fitting method relies on matching the strong feature, especially the 4000\AA\ break (3750-3950\AA). Removing the band with wavelengths greater than 3750\AA\ will result in a poor estimation in the corresponding redshift range, which can conveniently obtained by $z_{\rm min} = \lambda_{\rm L}/3950{\rm \AA} - 1$ and $z_{\rm max} = \lambda_{\rm R}/3750{\rm \AA} - 1$. Here, $\lambda_{\rm L}$ and $\lambda_{\rm R}$ are the wavelengths at both ends of the gap created by removing a band. In Figure~\ref{fig:band}, we find that the $F450M$, $F500M$, $F630M$ and $F763M$ bands mainly affect the fittings at $z\simeq0-0.3$, $0.2-0.6$, $0.3-1.0$ and $0.6-1.2$, respectively. By checking these areas on the $z_{\rm input}$ vs. $z_{\rm output}$ maps, we can find that the dots do not follow tightly the line $z_{\rm output}=z_{\rm input}$, but spread around it. Since the missing band affects the photo-$z$ accuracy within the corresponding redshift range, the redshift distribution will indirectly affect the photo-$z$ accuracy. For example, we have $\sigma_z^*=0.021, 0.024$ and $f_{\rm c}^*\sim6.52\%, 3.64\%$ if removing the $F630M$ and $F763M$ band, respectively, that can decrease the photo-$z$ fitting accuracy dramatically. An important reason is that $F630M$ and $F763M$ bands cover the peak (at $z\sim0.7$) of the galaxy redshift distribution of our catalog.

We can find that there are some catastrophic redshift dots on the left and bottom of $z_{\rm input}$ vs. $z_{\rm output}$ maps, which are due to the misidentification of spectral features at short wavelength as the features at long wavelength. According to \cite{Cao18}, the addition of ultraviolet and infrared bands can be effectively reduce the rate of catastrophic redshifts. As shown in Table~\ref{tab:band}, the values of $\sigma_z$ and $f_{\rm c}$ are similar for different filter groups, especially for $\sigma_z$. The main  reason for this phenomenon is that the SC filters fill the gaps between the MCI medium-band filters.

\subsection{Dependency of photo-$z$ accuracy on filter parameters}
\label{sect:para}

\begin{figure}
\centering
\includegraphics[width=0.496\columnwidth]{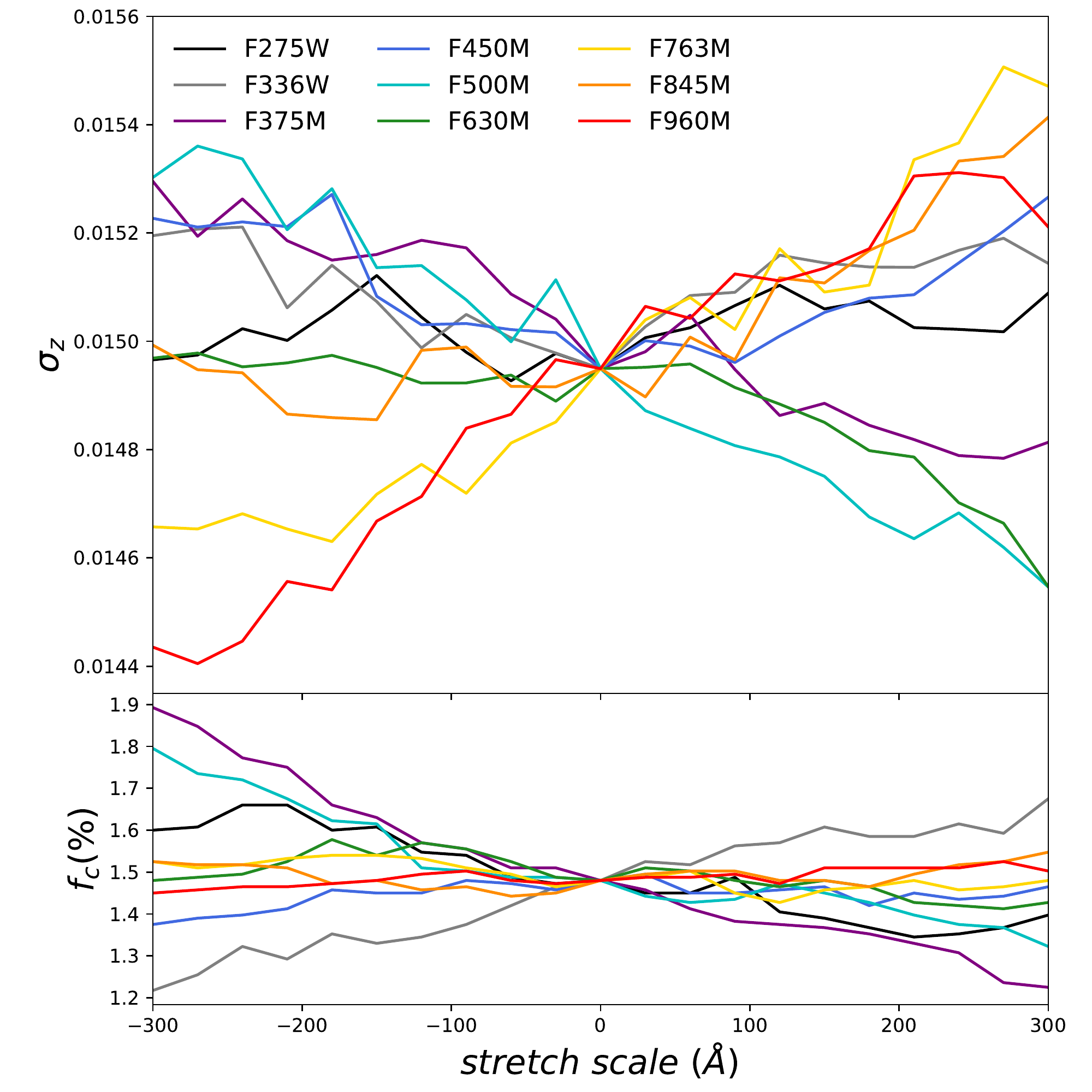}
\includegraphics[width=0.496\columnwidth]{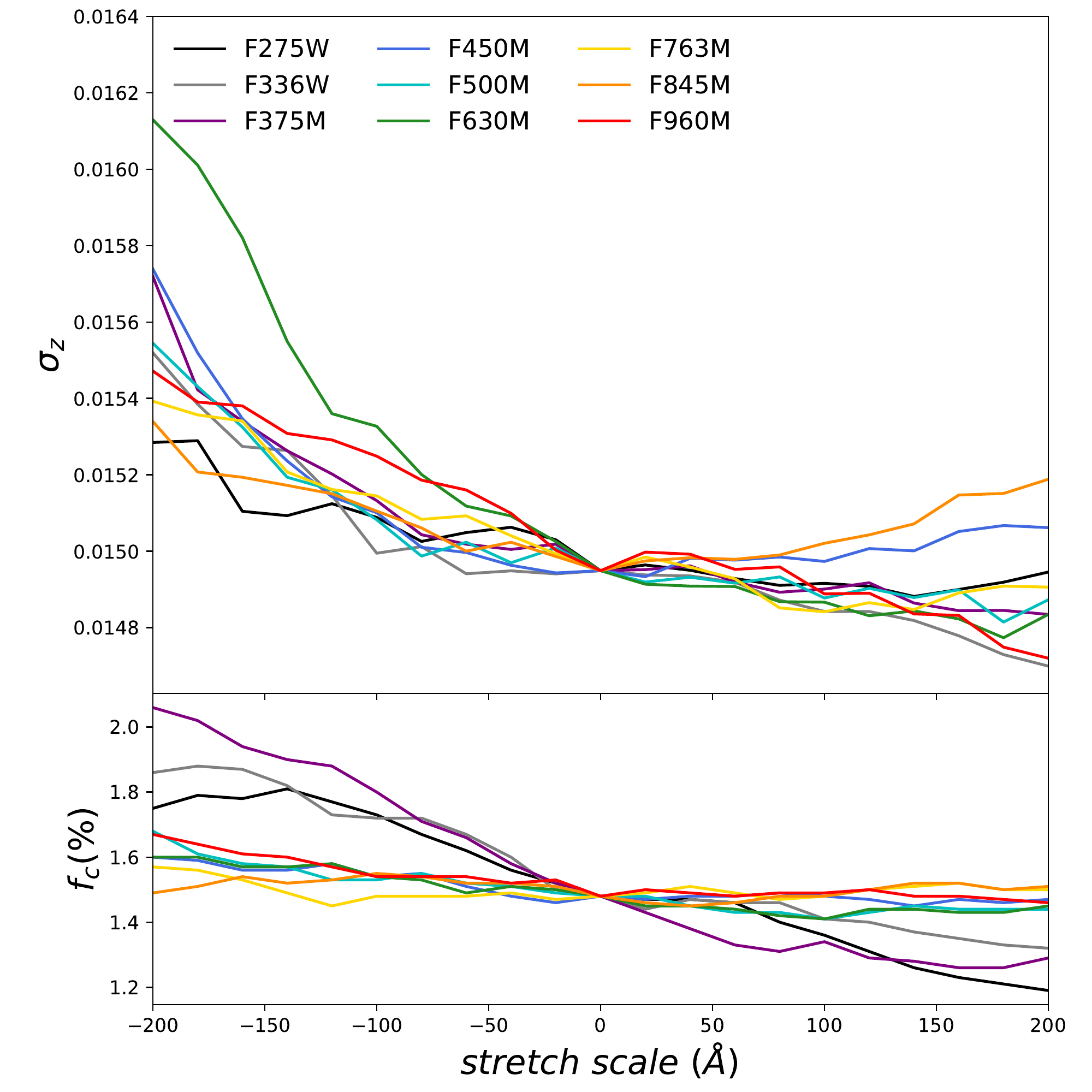}
\caption{$Left:$ the $\sigma_z$ and $f_{\rm c}$ as functions of the parameter $\lambda_{\rm c}^x$. $Right:$ the results as functions of the width stretch scale $\Delta \lambda_{\rm s}^x$. Here the position equal to $0$ represents the photo-$z$ fitting result using the original filter, which are shown in Figure~\ref{fig:filters}.}
\label{fig:para} 
\end{figure}

Here, we investigate the effect of filter parameters on photo-$z$ accuracy, and study the dependency of photo-z accuracy on these parameters to help testing and designing the MCI medium-band filters. We focus on two main filter parameters: the central wavelength position $\lambda_{\rm c}^x$ and the band wavelength coverage $\Delta \lambda^x$ for $x$ band. We first change the value of $\lambda_{\rm c}^x$ or $\Delta \lambda^x$ for one MCI medium-band filter, generate the mock data using the modified MCI medium-band filters and SC filters, and calculate the $\sigma_z$ and $f_{\rm c}$. Then we repeated the above process with different values of $\lambda_{\rm c}^x$ or $\Delta \lambda^x$. Finally we take the $\sigma_z$ and $f_{\rm c}$ as functions of the parameters to investigate the effect on photo-$z$ accuracy, and show the curves of $\sigma_z$ and $f_{\rm c}$ as functions of $\lambda_{\rm c}^x$ and $\Delta \lambda^x$ in Figure~\ref{fig:para}.

In the left panel of Figure~\ref{fig:para}, we show the $\sigma_z$ and $f_{\rm c}$ as a function of the shift scale $\Delta \lambda_{\rm c}^x$ for the MCI medium-band filter $x$. Here we use a shift scale in $\rm \AA$ to denote the positions of central wavelength, and the data at shift scale $\Delta \lambda_c^x=0$ represents the result for the original position case, which are given in Figure~\ref{fig:filters} and Table~\ref{tab:filters}. Without changing the shape of the filter, we will gradually shift the central wavelength position of filters from -300 to +300$\rm \AA$ of the original positions, and the scale of each shift is 30$\rm \AA$. We can find that $\sigma_z$ and $f_{c}$ is close to the minimum values at the original position for most curves. However, a bluer $F336W$, $F763W$ and $F960M$ band or redder $F275W$, $F375M$, $F500M$ and $F630M$ bands may be better.
As the discussion in the Section~\ref{sect:band}, changing the near-ultraviolet bands $F275W$, $F336W$ and $F375M$ mainly improves the number of catastrophic redshift in photo-$z$ fitting. The bluer $F960M$ can significantly improves the deviation $\sigma_z$ of photo-$z$ estimation.
This is because moving the $F960M$ band towards the blue end can help improving the efficiency of the filter under the influence of the detector quantum efficiency curve.

In the right panel of Figure~\ref{fig:para}, we show the $\sigma_z$ and $f_{\rm c}$ as a function of the width stretch scale $\Delta \lambda^x_{\rm s}$, which is adopted to adjust the FWHM by $\Delta \lambda=\Delta \lambda_{\rm ori} + \Delta \lambda_{\rm s} $. Here $\Delta \lambda_{\rm ori}$ is the FWHM of the original filter shown in Table~\ref{tab:filters}. We can find that the wider filter is helpful to improve the accuracy of photo-$z$ fitting, and most $\sigma_z$ and $f_{c}$ of the original filters are around the minimum values.
According to the curve of $f_{\rm c}$, only increasing the width of the near-ultraviolet filters can be helpful to pin down the number of catastrophic redshift.
Due to limited manufacturing technology, the efficiencies of the near-ultraviolet filters are lower than other filters. The simulation results show that we can reduce the effect of the inefficient filter on the photo-$z$ fitting by increasing the width and exposure time. A wider $F630M$ and $F960M$ bands can be helpful to reduce the deviation. 
But the improvements of $\sigma_z$ are not much (less than 0.001), and this indicates that the original filters are proper for the photo-$z$ calibration.

\section{Summary}
\label{sect:sum}

In this work, we investigate the accuracy of photo-$z$ measurement with CSST-MCI medium-band filters, which cover a large wavelength range from near-ultraviolet to near-infrared bands. 
We select high-quality galaxies with SNR $\ge$10 in $g$ or $i$ band from the COSMOS08 catalog, and use these galaxy templates generate the mock catalog, which has peaks of the magnitude distribution and redshift distribution at $m_{i^+}\sim24.3$ and $z\sim 0.7$, respectively. Then we randomly select 10,000 galaxies from the high-quality samples, and generate the mock data and errors based on CSST instrument parameters. Then we measure the photo-$z$ and analyze the results with two parameters $\sigma_z$ and $f_c$.

First, we build the SED template library using 31 extend SEDs, which include elliptical, spiral and star-forming galaxy templates. We simulate the extinction of dust from a galaxy itself for a restframe SED, and then shift the processed SED to the given redshift from the mock catalog. After considering the absorption of the IGM, we calculate the flux by convolving the final SED with the filter transmissions, and add Gaussian noise to the final mock data. The uncertainties of data are obtained by calculating errors from the instrumental noise, sky background and systematic errors. We adopt the modified LePhare code to fit the photo-$z$ using both MCI and SC data, and we find that the results can achieve $\sigma_z\sim0.015$ and $f_{\rm c}\sim1.5\%$.

Next, we investigate the effect of each MCI medium-band filter passband on photo-$z$ accuracy by removing it. The mock data are generated after removing one band each time, and then we perform the  fitting process and analyze the photo-$z$ accuracy. We find that the accuracy of the SED template fitting method depends on the recognition of 4000\AA\ break, and the bands with wavelengths greater than 3750\AA\ can significantly affect the photo-$z$ fitting accuracy. Due to the influence of redshift distribution, the $F630M$ and $F763M$ bands have the largest effects on $\sigma_z$ and $f_{\rm c}$. On the other hand, the near-ultraviolet bands have relatively small influence on the accuracy of photo-$z$.

Finally, we investigate the dependency of photo-z accuracy on the design parameters of the MCI medium-band filters. We focus on central wavelength position $\lambda_c$ and wavelength coverage $\Delta$ for each band, and estimate the accuracy for the certain filter parameter ranges. We find that the $\sigma_z$ and $f_c$ are always less than 0.016 and 2\% (around 0.015 and 1.5\%), respectively, in the filter parameter ranges we explore. We find that the bluer $F336W$ and $F960M$ bands or redder $F275W$, $F375M$, $F500M$ and $F630M$ bands are more advantageous to the photo-$z$ estimation, but the improvements of $\sigma_a$ are less than 0.001. This indicates that the original MCI medium-band filters are proper for photo-z calibration.

\normalem
\begin{acknowledgements}

Y.C. and Y.G. thank Xianmin Meng for helpful discussion. Y.C. and Y.G. acknowledge the support of NSFC-11822305, NSFC-11773031, NSFC-11633004, MOST-2018YFE0120800, MOST-2020SKA0110402, and CAS Interdisciplinary Innovation Team. This work is also supported by the science research grants from the China Manned Space Project with NO.CMS-CSST-2021-B01 and CMS-CSST-2021-A01. Z.Y.Z. acknowledges support by the National Science Foundation of China (11773051, 12022303) and the CAS Pioneer Hundred Talents Program.

 \end{acknowledgements}

\bibliographystyle{raa}
\bibliography{MCI.bib}

\clearpage

\end{document}